\documentclass[twocolumn,aps,prd,longbibliography]{revtex4}
\usepackage{dcolumn}
\usepackage{enumerate}

\usepackage{amsmath}
\usepackage{amssymb}
\usepackage{graphicx}
\usepackage{graphics}

\begin{document}

\title{Two types of axisymmetric helical magnetorotational instability
in rotating flows with positive shear}

\author{George Mamatsashvili}
\email{george.mamatsashvili@nbi.ku.dk}

\affiliation{Niels Bohr International Academy, Niels Bohr Institute,
Blegdamsvej 17, 2100 Copenhagen, Denmark}

\affiliation{Helmholtz-Zentrum Dresden-Rossendorf, Bautzner Landstr.
400, D-01328 Dresden, Germany}

\author{Frank Stefani}
\affiliation{Helmholtz-Zentrum Dresden-Rossendorf, Bautzner Landstr.
400, D-01328 Dresden, Germany}

\author{Rainer Hollerbach}
\affiliation{Department of Applied Mathematics,
University of Leeds, Leeds LS2 9JT, U.K.}

\author{G\"unther R\"udiger}
\affiliation{Leibniz-Institut f\"ur Astrophysik Potsdam,
An der Sternwarte 16, D-14482 Potsdam, Germany}

\date{\today}

\begin{abstract}
We reveal and investigate a new type of linear axisymmetric helical magnetorotational 
instability which is capable of destabilizing viscous and resistive rotational flows with radially 
increasing angular velocity, or positive shear. This instability is double-diffusive by nature and is 
different from the more familiar helical magnetorotational instability, operating at positive shear 
above the Liu limit, in that it works instead for a wide range of the positive shear when ${\rm 
(i)}$ a combination of axial/poloidal and azimuthal/toroidal magnetic fields is applied and ${\rm (ii)}$
the magnetic Prandtl number is not too close to unity. We study this instability first with radially 
local WKB analysis, deriving the scaling properties of its growth rate with respect to Hartmann, 
Reynolds and magnetic Prandtl numbers. Then we confirm its existence using a global stability 
analysis of the magnetized flow confined between two rotating coaxial cylinders with purely 
conducting or insulating boundaries and compare the results with those of the local 
analysis. From an experimental point of view, we also demonstrate the presence of the new 
instability in a magnetized viscous and resistive Taylor-Couette flow with positive shear for such 
values of the flow parameters, which can be realized in upcoming experiments at the DRESDYN 
facility. Finally, this instability might have implications for the dynamics of the equatorial parts of 
the solar tachocline and dynamo action there, since the above two necessary conditions for the 
instability to take place are satisfied in this region. Our global stability calculations for 
the tachocline-like configuration, representing a thin rotating cylindrical layer with the appropriate 
boundary conditions -- conducting inner and insulating outer cylinders -- and the values of the flow 
parameters, indicate that it can indeed arise in this case with a characteristic growth time 
comparable to the solar cycle period.
\end{abstract}


\maketitle

\section{Introduction}

According to Rayleigh's criterion \cite{Rayleigh_1917}, rotating flows of ideal fluids with radially 
increasing specific angular momentum are linearly stable. This result has severe astrophysical 
consequences, implying hydrodynamic stability of Keplerian rotation in accretion disks. Nowadays, 
the magnetorotational instability (MRI) \cite{Velikhov_1959,Chandra_1960,Balbus_Hawley1991} is 
considered to be the most likely destabilizing mechanism for these disks, driving radially outward 
transport of angular momentum and inward accretion of mass.

The standard MRI (SMRI, with a purely axial/poloidal magnetic field, 
\cite{Velikhov_1959,Chandra_1960,Balbus_Hawley1991}), as well  as the non-axisymmetric 
azimuthal MRI (AMRI, with a purely azimuthal/toroidal magnetic field, \cite{Hollerbach_etal2010}) 
and the axisymmetric helical MRI (HMRI, with combined axial and azimuthal magnetic fields, 
\cite{Hollerbach_Ruediger2005}) have all been extensively studied theoretically (see a recent 
review \cite{Ruediger_etal2018a} and references therein). The inductionless forms of AMRI and 
HMRI have also been obtained in liquid metal experiments 
\cite{Seilmayer_etal2014,Stefani_etal2006,Stefani_etal2009}, while unambiguous experimental 
evidence for inductive SMRI remains elusive, despite promising first results 
\cite{Sisan_etal2004,Nornberg_etal2010}.

In contrast to Keplerian-like rotation with increasing angular momentum but decreasing angular 
velocity, much less attention is usually devoted to flows with increasing angular velocity. Until 
recently, such flows have been believed to be strongly stable, even under magnetic fields. 
However, for very high enough Reynolds numbers $Re\sim 10^7$, they can yield non-axisymmetric 
linear instability \cite{Deguchi2017}. Apart from this hydrodynamic instability, there is also a 
special type of AMRI operating in flows with much lower Reynolds number but sufficiently strong 
positive shear \cite{Stefani_Kirillov2015,Ruediger_etal2016,Ruediger_etal2018b}. This restriction to 
strong shear makes, however, this so-called {\it Super-AMRI} astrophysically less significant. One 
of few positive shear regions is a portion of the solar tachocline extending $\pm 30^{\circ}$ about 
the Sun's equator. Even there, the shear measured in terms of Rossby number $Ro= 
r(2\Omega)^{-1} d\Omega/dr$ is only around 0.7 \cite{Parfrey_Menou2007,Tobias_etal2007}, 
much less than the so-called {\it upper Liu limit} (ULL) $Ro_{ULL}=2(1+\sqrt{2})\approx  4.83$ 
\cite{Liu_etal2006} required for Super-AMRI. Another astrophysical system in which positive 
shear is expected is the boundary layer between an accretion disk and its host star
\cite{Belyaev_etal2012, Pessah_Chan2012}.

Given a general similarity between AMRI and HMRI and the universal nature of the Liu limits 
\cite{Kirillov_Stefani2012,Kirillov_etal2014}, one might expect a similar result to hold also for {\it 
Super-HMRI}. However, as we report in this paper, there exists a new type of axisymmetric HMRI, 
which we refer to as type 2 Super-HMRI, that operates in positive shear flows with arbitrary 
steepness, whereas the more familiar HMRI operating only at high enough positive shear above the 
Liu limit, $Ro > Ro_{ULL}$, is labelled type 1 Super-HMRI. The only requirements are {\rm (i)} the 
presence of both axial and azimuthal magnetic field components and {\rm (ii)} magnetic Prandtl 
number is neither zero (the inductionless limit) nor too close to unity. These conditions are 
indeed satisfied in the solar tachocline, where this new instability can possibly play an 
important role in its dynamics and magnetic activity. Although this requires a 
detailed separate study and is out of scope of the present paper, we have also done 
calculations at the end of this paper, showing the possibility of occurrence of this 
instability for the tachocline-like configuration and parameters, but still remaining in the framework 
of cylindrical flow. The resulting growth time (inverse of the exponential growth rate) of the most 
unstable mode in fact turns out to be comparable to the solar cycle period.  

In this paper, we carry out a linear stability analysis of a magnetic rotational flow in cylindrical 
geometry mainly using the Wentzel-Kramers-Brillouin (WKB) short-wavelength formulation of the 
underlying magnetohydrodynamics (MHD) problem \cite{Kirillov_Stefani2010,Kirillov_etal2014}, 
which is especially useful for understanding the basic features and scaling properties of the new 
instability. This local analysis is then complemented by global, radially one-dimensional (1D) 
calculations of the corresponding unstable eigenmodes with the primary aim to demonstrate the 
existence of this new version of Super-HMRI beyond the local WKB approximation as well as to 
draw a comparison with the results obtained using this approximation. A more comprehensive 
global linear analysis exploring parameter space, and subsequently nonlinear analysis of this 
double-diffusive type 2 Super-HMRI at positive shear will be presented elsewhere. 

The paper is organized as follows. Main equations and the formulation of a problem are given in 
Sec. II.  The local WKB analysis of the instability is presented in Sec. III. The global stability analysis 
of a differentialy rotating flow between two coaxial cylinders at positive shear both in the narrow 
and wide gap cases as well as a comparison with the results of the local analysis are presented in 
Sec. IV. A summary and discussion on the relevance of this new version of Super-HMRI to the 
solar tachocline are given in Sec. V.

\section{Main equations}

The motion of an incompressible conducting medium with constant viscosity $\nu$ and ohmic 
resistivity $\eta$ is governed by the equations of non-ideal MHD
\begin{equation}
\frac{\partial {\bf U}}{\partial t}+({\bf U}\cdot \nabla) {\bf U}=-\frac{1}{\rho}\nabla \left(P+\frac{{\bf 
B}^2}{2\mu_0}\right)+\frac{({\bf B}\cdot\nabla) {\bf B}}{\mu_0 \rho} + \nu\nabla^2 {\bf U},
\end{equation}
\begin{equation}
\frac{\partial {\bf B}}{\partial t}=\nabla\times \left( {\bf U}\times {\bf B}\right)+\eta\nabla^2{\bf B},
\end{equation}
\begin{equation}
\nabla\cdot {\bf U}=0,~~~\nabla\cdot {\bf B}=0.
\end{equation}
where $\rho$ is the constant density, ${\bf U}$ is the velocity, $P$ is the thermal pressure,  
${\bf B}$ is the magnetic field and $\mu_0$ is the magnetic permeability of vacuum.

Consider a flow between two coaxial cylinders at inner, $r_i$, and outer, $r_o$, radii,  rotating, 
respectively, with angular velocities $\Omega_i$ and $\Omega_o$ in the cylindrical coordinates 
$(r,\phi,z)$. Since we are primarily interested in the flow stability in the case of positive shear, or 
so-called ``super-rotation'' \cite{Ruediger_etal2016,Ruediger_etal2018b}, the inner cylinder is 
assumed to rotate slower than the outer one, $\Omega_i<\Omega_o$, inducing an azimuthal 
nonuniform flow ${\bf U}_0=(0,r\Omega(r),0)$ between the cylinders with radially increasing 
angular velocity, $d\Omega/dr>0$, and hence positive Rossby number, $Ro>0$. The pressure 
associated with this base flow and maintaining its rotation is denoted as $P_0$. The imposed 
background helical magnetic field ${\bf B}_0=(0, B_{0\phi}(r), B_{0z})$ consists of a radially 
varying, current-free azimuthal component, $B_{0\phi}(r) = \beta B_{0z}r_o/r$, and a constant axial 
component, $B_{0z}$, where the constant parameter $\beta$ characterizes field's helicity. 

We investigate the linear stability of this equilibrium against small {\it axisymmetric} ($\partial 
/\partial \phi=0$) perturbations, ${\bf u}={\bf U}-{\bf U}_0$, $p=P-P_0$, ${\bf b}={\bf B}-{\bf 
B}_0$, which are all functions of $r$ and depend on time $t$ and axial/vertical $z$-coordinate via 
$\propto \exp(\gamma t+ {\rm i}k_z z)$, where $\gamma$ is the (complex) eigenvalue and $k_z$ is 
the axial wavenumber. There is instability in the flow, if the real part of any eigenvalue, or  
growth rate is positive, ${\rm Re}(\gamma)>0$, for any of the eigennmodes. In such cases, for a 
given set of parameters, we always select out the mode with the largest growth rate from a 
corresponding eigenvalue spectrum.

\section{WKB analysis}

In this section, we use a radially local WKB approximation, where the radial dependence of the 
perturbations is assumed to be of the form $\propto\exp({\rm i}k_r r)$ with $k_r$ being the radial 
wavenumber. The resulting dispersion relation, which follows from Eqs. (1)-(3) after linearizing and 
substituting the above exponential form of the perturbations, is represented by the fourth-order 
polynomial  
\cite{Kirillov_Stefani2010,Kirillov_etal2014}:
\begin{equation}
\gamma^4+a_1\gamma^3+a_2\gamma^2+(a_3+{\rm i}b_3)\gamma+a_4+{\rm i}b_4=0,
\end{equation}
with the real coefficients
\begin{equation*}
a_1=2\frac{k^2}{Re}\left(1+\frac{1}{Pm}\right),
\end{equation*}
\begin{multline*}
a_2=4\alpha^2(1+Ro)+2(k_z^2+2\alpha^2\beta^2)\frac{Ha^2}{Re^2Pm}\\
+\frac{k^4}{Re^2}\left(1+\frac{4}{Pm}+\frac{1}{Pm^2}\right),
\end{multline*}
\begin{multline*}
a_3=8(1+Ro)\alpha^2\frac{k^2}{RePm}\\+ 
2[k^4+(k_z^2+2\alpha^2\beta^2)Ha^2]\frac{k^2}{Re^3Pm}\left(1+\frac{1}{Pm}\right)
\end{multline*}
\begin{equation*}
b_3=-8\alpha^2\beta k_z\frac{Ha^2}{Re^2Pm},
\end{equation*}
\begin{multline*}
a_4=4\alpha^2 \frac{k^4}{Pm^2}\left[(1+Ro)\frac{1}{Re^2}+\beta^2\frac{Ha^2}{Re^4}\right]
\\+4\alpha^2k_z^2Ro\frac{Ha^2}{Re^2Pm}+\left(k_z^2Ha^2+k^4\right)^2\frac{1}{Re^4Pm^2},
\end{multline*}
\begin{equation*}
b_4=4\beta k_z^3\left[Ro\left(1-\frac{1}{Pm}\right)-\frac{2}{Pm}\right]
\frac{Ha^2}{Re^3Pm}.
\end{equation*}
Henceforth $\gamma$ is normalized by the outer cylinder's angular velocity 
$\Omega_o$, and the wavenumbers by its inverse radius, $r_o^{-1}$. Other nondimensional 
parameters are: $\alpha=k_z/k$, where $k=(k_r^2+k_z^2)^{1/2}$ is the total wavenumber; the 
Reynolds number $Re=\Omega_o r_o^2/\nu$, the magnetic Reynolds number $Rm=\Omega_o 
r_o^2/\eta$, and their ratio, the magnetic Prandtl number $Pm=\nu/\eta=Rm/Re$; the Hartmann 
number $Ha=B_{0z}r_o/(\mu_0\rho\nu \eta)^{1/2}$ that measures the strength of the imposed axial 
magnetic field. Another quantity characterizing the field is Lundquist number $S=HaPm^{1/2}$, 
which, like $Rm$, does not involve viscosity. Since we focus on positive Rossby 
numbers, $Ro>0$, or positive shear, the flow is generally stable both hydrodynamically, according 
to Rayleigh's criterion (but see Ref. \cite{Deguchi2017}), as well as against SMRI with a purely axial 
field ($\beta=0$) \cite{Ji_etal2001,Goodman_Ji2002,Kirillov_Stefani2010}. 

In the inductionless limit, $Pm\rightarrow 0$, the roots of Eq.\ (4) can be found analytically 
\cite{Liu_etal2006,Kirillov_Stefani2010,Priede2011,Kirillov_etal2014,Mamatsashvili_Stefani2016}. 
For positive and relatively large $Ro>Ro_{ULL}$, one of the roots always has a positive real part, 
implying instability with the growth rate
\begin{multline}
{\rm Re}(\gamma)=\sqrt{2X+2\sqrt{X^2+Y^2}}\\
-(k_z^2+2\alpha^2\beta^2)\frac{Ha^2}{k^2Re}-\frac{k^2}{Re},
\end{multline}
where
\begin{equation*}
X=\alpha^2\beta^2(\alpha^2\beta^2+k_z^2)\frac{Ha^4}{Re^2k^4}-\alpha^2(1+Ro),
\end{equation*}
\begin{equation*}
Y=\beta \alpha^2 k_z(2+Ro)\frac{Ha^2}{k^2Re},
\end{equation*}
which we call type 1 Super-HMRI. \emph{Our main goal though is to reveal that apart from this type 
1 Super-HMRI at large positive shear, Eq.\ (4) also yields a completely new type of 
dissipation-induced double-diffusive instability at finite $Pm$, which we call type 2 Super-HMRI}.

Regarding the dependence on $\beta$ parameter in Eq.\ (4), it is readily seen that, as 
long as $\beta \neq 0$, it enters the coefficients of these dispersion relations through the 
re-scaled wavenumbers, Hartmann, Lundquist and Reynolds numbers, $k_z^{\ast}\equiv 
k_z/\beta$, $k^{\ast}\equiv k/\beta$, $Ha^{\ast}\equiv Ha/\beta$, $S^{\ast}\equiv S/\beta$, 
$Re^{\ast}\equiv Re/\beta^2$, $Rm^{\ast}\equiv Rm/\beta^2$, in terms of which we carry out the 
following WKB analysis. It is easy to check that $\beta$ disappears in the polynomial Eq. (4) 
after substituting these re-scaled parameters (denoted with asterisks) in its coefficients.  

Figure 1(a) shows the growth rate, ${\rm Re}(\gamma)$, as a function of the re-scaled axial 
wavenumber, as determined from a numerical solution of Eq. (4) at finite but very small 
$Pm=10^{-6}$, together with solution (5) in the inductionless limit, for fixed $Ha^{\ast}$ and 
$Re^{\ast}$. For the Rossby number we take the values lower, $Ro=1.5,2$, and higher, $Ro=6$, 
than $Ro_{ULL}$. Two distinct instability regimes are clearly seen in this figure. Type 2 Super-HMRI 
is concentrated at small $k_z^{\ast}$ and exists at finite $Pm$ both for $Ro<Ro_{ULL}$ and 
$Ro>Ro_{ULL}$, i.e., it is insensitive to the upper Liu limit, but disappears for $Pm\rightarrow0$ 
at fixed Hartmann and Reynolds numbers. By contrast, type 1 Super-HMRI, concentrated at larger 
$k_z^{\ast}$, exists only for $Ro>Ro_{ULL}$, and approaches the inductionless solution as 
$Pm\rightarrow 0$. This latter branch is basically an extension of the more familiar HMRI operating 
at negative shear, which in the inductionless limit also satisfies Eq. (5), but at $Ro<Ro_{LLL}$, 
where $Ro_{LLL}=2(1-\sqrt{2})\approx -0.83$ is the lower Liu limit \cite{Liu_etal2006, 
Kirillov_Stefani2010, Priede2011}. 

At large $Pm\gg 1$, type 1 Super-HMRI disappears and there remains only type 2 
Super-HMRI, as shown in Fig. 2(a). The corresponding dispersion curves as a function of axial 
wavenumber have a shape similar to those at small $Pm$ in Fig. 1(a), but now the instability occurs 
at order of magnitude larger $k_z^{\ast}$ and several orders of magnitude smaller $Ha^{\ast}$ and 
$Re^{\ast}$ at the same values of $Ro$ adopted in these figures.

Thus, type 2 Super-HMRI represents a new, dissipation-induced instability mode at positive shear, 
which appears to require the presence of \emph{both} finite viscosity and resistivity. As we will see 
below though, it does not operate in the immediate vicinity of $Pm=1$, that is, it is double-diffusive 
in nature, operating for both small and large $Pm$, but not for $Pm=O(1)$. Just as all previous MRI 
variants, this one also derives energy solely from the shear, since the imposed magnetic field is 
current-free, thereby eliminating current-driven instabilities, such as the Tayler instability. Energy is 
drawn from the background flow $r\Omega(r)$ to the growing perturbations due to the coupling 
between meridional circulation and azimuthal field perturbations brought about by the imposed 
azimuthal field, a mechanism also underlying HMRI at negative shear 
\cite{Hollerbach_Ruediger2005,Priede_etal2007}.

Our main goal is to describe the properties of this new type 2 Super-HMRI. Type 1 Super-HMRI, 
existing only for $Ro>Ro_{ULL}$ and persisting even in the inductionless limit $Pm\rightarrow0$ 
\cite{Liu_etal2006,Kirillov_Stefani2010,Kirillov_etal2014,Mamatsashvili_Stefani2016}, is also 
relatively new and interesting in its own right, but will not be considered here further.

Like normal HMRI at negative shear, type 2 Super-HMRI is an overstability, that is, its growth 
rate comes with an associated non-zero imaginary part, $\omega={\rm Im}(\gamma)$, which is 
the frequency of temporal oscillations of the solution at a given coordinate and, together with axial 
wavenumber, defines its propagation speed. Figure 3 shows these frequencies as a function of 
$k_z^{\ast}$, corresponding to the growth rates plotted in Figs.\ 1(a) and 2(a). They monotonically 
increase with $k_z^{\ast}$ by absolute value, but are positive at small $Pm$ and negative at large 
$Pm$, implying opposite propagation directions of the wave patterns at these magnetic Prandtl 
numbers. Also, $\omega$ remains smaller than the frequency of inertial oscillations, 
$\omega_{io}=2\alpha(1+Ro)^{1/2}$, and tend to the latter only at small $Pm$ as the solution 
changes from type 2 to type 1 Super-HMRI with increasing $k_z^{\ast}$ and do not change 
afterwards. This reflects the fact that type 1 Super-HMRI represents weakly 
destabilized inertial oscillations, like the normal HMRI at negative shear \cite{Liu_etal2006}.

\begin{figure}
\includegraphics[width=\columnwidth]{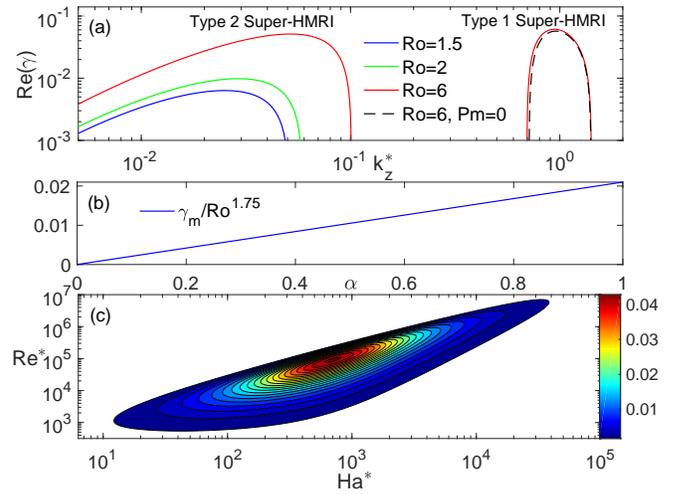}
\caption{Panel (a) shows the growth rate ${\rm Re}(\gamma)$ vs. $k_z^{\ast}$ at fixed 
$Ha^{\ast}=90$, $Re^{\ast}=8\times10^3$, $\alpha=0.71$ (i.e., $k_r^{\ast}=k_z^{\ast}$) and 
$Pm=10^{-6}$ for different $Ro=1.5~(blue),~2~(green),~6~(red)$. New type 2 Super-HMRI branch 
exists at smaller $k_z^{\ast}$ and finite $Pm$, for all three $Ro$ values. By contrast, type 1 
Super-HMRI branch at larger $k_z^{\ast}$ appears only for $Ro=6>Ro_{ULL}$ from these there 
values of the Rossby number, but persists also in the inductionless limit (Eq.\ 5, dashed-black line). 
For the same $Pm$, panel (b) {\normalsize {\large }}shows the growth rate of type 2 Super-HMRI, 
maximized over a set of the parameters ($k_z^{\ast}$, $Ha^{\ast}$, $Re^{\ast}$) and normalized 
by $Ro^{1.75}$, vs. $\alpha$, while panel (c) shows the growth rate, maximized over $k_z^{\ast}$ 
and $\alpha$, as a function of  $Ha^{\ast}$ and $Re^{\ast}$ at $Ro=1.5$ and the same 
$Pm=10^{-6}$.}
\end{figure}

\begin{figure}
\includegraphics[width=\columnwidth]{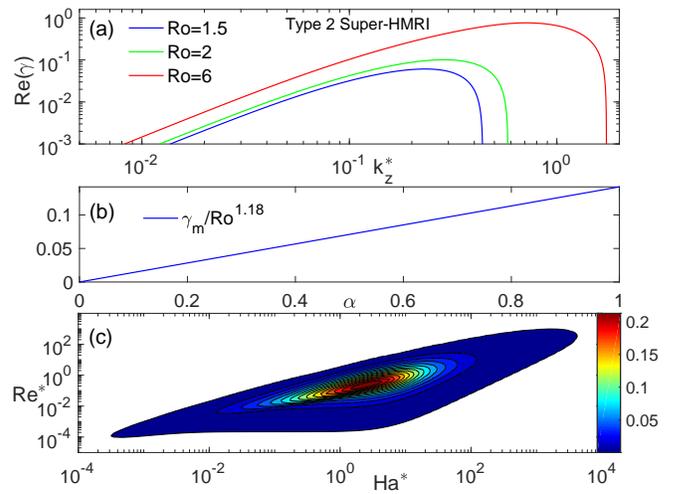}
\caption{Same as in Fig.\ 1, but at $Ha^{\ast}=5$, $Re^{\ast}=0.1$, $\alpha=0.71$ in panel (a) and 
$Pm=100$ in all panels. New type 2 Super-HMRI branch exists at higher $k_z^{\ast}$ than those at 
small $Pm$, while type 1 Super-HMRI branch is absent. In panel (b), the maximum growth rate now 
exhibits the scaling with Rossby number, $Ro^{1.18}$, different from that at small $Pm$. In panel 
(c), the maximum growth occurs now at orders of magnitude smaller $Ha_m^{\ast}$ and 
$Re_m^{\ast}$ than those at small $Pm$ in Fig.\ 1(c) at the same $Ro=1.5$.}
\end{figure}

\begin{figure}
\includegraphics[width=\columnwidth]{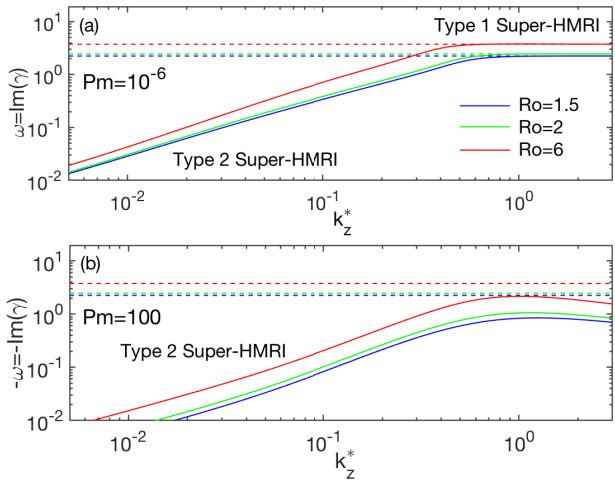}
\caption{Imaginary part of the eigenvalues, that is, frequency $\omega={\rm Im}(\gamma)$, at 
small $Pm=10^{-6}$ (a) and large $Pm=100$ (b) corresponding, respectively, to the growth rates 
shown in Figs. 1(a) and 2(a) at the same $Ro=1.5~(blue)$, $Ro=2~(green)$, $Ro=~6 (red)$ 
and $\alpha=0.71$. Dashed lines correspond to the frequency of inertial oscillations, 
$\omega_{io}=2\alpha(1+Ro)^{1/2}$, for these parameters. Note that at small $Pm$ 
the frequencies are positive, while at large $Pm$ are negative ($-{\rm Im}(\omega)$ is plotted in 
panel b), implying opposite propagation directions of the wave patterns at these magnetic Prandtl 
numbers.}
\end{figure}

To explore the behavior of type 2 Super-HMRI further, we first vary $\alpha$ as well as the 
re-scaled Hartmann and Reynolds numbers. The growth rate, maximized over the last two numbers 
and $k_z^{\ast}$, increases linearly with $\alpha$ and scales as $\propto Ro^{1.75}$ at small 
$Pm=10^{-6}$ (Fig.\ 1(b)) and as $\propto Ro^{1.18}$ at large $Pm=100$ (Fig.\ 2(b)), while its 
dependence on $Ha^{\ast}$ and $Re^{\ast}$, when maximized over $k_z^{\ast}$ and $\alpha$, is 
shown in Fig.\ 1(c) at $Pm=10^{-6}$ and in Fig.\ 2(c) at $Pm=100$ with $Ro=1.5<Ro_{ULL}$ 
(when type 1 Super-HMRI is absent) in both cases. The most unstable region is quite localized, 
with the growth rate decreasing for both small and large $Ha^{\ast}$ and $Re^{\ast}$, implying 
that this instability relies on finite viscosity and resistivity, i.e., it is indeed of double-diffusive type. 
The overall shape of the unstable area in $(Ha^{\ast},Re^{\ast})$-plane does not change 
qualitatively at other $Pm$ and $Ro$; the unstable region always remains localized and shifts to 
larger $Ha^{\ast}$ and $Re^{\ast}$ with decreasing $Pm$. In particular, the maximum growth rate, 
$\gamma_m$, occurs for $(Ha_m^{\ast}, Re_m^{\ast})\approx (700,~9\times10^4)$ when $Pm$ is 
small (Fig.\ 1(c)), but for orders of magnitude smaller $(Ha_m^{\ast}, Re_m^{\ast})=(2.54,~0.23)$ 
when $Pm$ is large (Fig.\ 2(c)) The actual values of the characteristic vertical wavenumber, 
Hartmann and Reynolds numbers for type 2 Super-HMRI at different $\beta$ are obtained by 
simply multiplying the values of re-scaled quantities $k_z^{\ast}, Ha^{\ast}, Re^{\ast}$ given in 
Figs.\ 1 and 2 (as well as those in Fig. 4 of the following analysis) by $\beta$ or
$\beta^2$, respectively. For example, the largest growth rate at $Pm=10^{-6}$ actually occur at 
$(Ha_m, Re_m)=\beta(Ha_m^{\ast}, Re_m^{\ast}\beta)\approx \beta(700,~9\times10^4\beta)$. 
These latter values can, in turn, be compared with the global stability analysis presented below as 
well as with magnetic Taylor-Couette (TC) flow experiments.

We next consider variation with $Pm$ at fixed $Ro=1.5<Ro_{ULL}$, so that type 1 Super-HMRI is 
excluded. Figure 4 shows the growth rate $\gamma_m$, maximized over a set of parameters 
$(\alpha,k_z^{\ast}, Ha^{\ast}, Re^{\ast})$ as well as the associated $Ha_m^{\ast}$ and 
$Re_m^{\ast}$, at which this maximum growth is achieved, as a function of $Pm$. It is seen that 
for $Pm\lesssim 10^{-2}$, the growth rate is practically constant, $\gamma_m=0.043$, 
while $Ha_m^{\ast}$ and $Re_m^{\ast}$ increase with decreasing $Pm$ as the power-laws: 
$Ha_m^{\ast}\propto Pm^{-1/2}$ and $Re_m^{\ast}\propto Pm^{-1}$. So, in this small-$Pm$ 
regime, type 2 Super-HMRI is more appropriately described in terms of Lundquist and magnetic 
Reynolds numbers, since these are $S_m^{\ast}=Ha_m^{\ast}Pm^{1/2}=0.7$,
$Rm_m^{\ast}=Re_m^{\ast}Pm=0.091$, and thus independent of $Pm$. These scalings with 
$S$ and $Rm$, being independent of $Pm$ at $Pm\rightarrow0$, are similar to those of SMRI, 
and imply that this new instability also does not exist in the inductionless limit, which would require 
$S^{\ast},Rm^{\ast}\rightarrow0$, and hence $S,Rm\rightarrow 0$, if $Ha^{\ast}$ and $Re^{\ast}$ 
are to remain finite.

With increasing $Pm$, beyond $Pm\sim 0.01$, $\gamma_m$ rapidly decreases, and eventually the 
instability disappears at the first critical value $Pm_{c1}=0.223$, with corresponding 
$(Ha_m^{\ast}, Re_m^{\ast})=(2.018,~0.071)$ and the re-scaled wavenumber 
$k_{zm}^{\ast}=2.4\times10^{-3}$. It reappears again for larger $Pm>1$ at the second critical value 
$Pm_{c2}=4.46$, with $(Ha_m^{\ast}, Re_m^{\ast})=(2,~0.046)$ and 
$k_{zm}^{\ast}=5\times10^{-3}$, comparable to those at $Pm_{c1}$. Further increasing $Pm$, for 
$Pm\gtrsim 10$, $\gamma_m$ eventually approaches a constant value 0.29. The corresponding 
re-scaled Hartmann and Reynolds numbers again follow the power-law scalings, now  
$Ha_m^{\ast}\propto Pm^{1/3}$ and $Re_m^{\ast}\propto Pm^{-1/4}$.

Note that while we have only presented the $Ro=1.5$ case in order to demonstrate the behavior of 
the instability with $Pm$, other values of Rossby number yield qualitatively similar behavior and 
scalings of $\gamma_m$, $Ha_m^{\ast}$, $Re_m^{\ast}$. Thus, the new type 2 Super-HMRI exists 
over a broad range of magnetic Prandtl numbers, provided that viscosity $\nu$ and resistivity 
$\eta$ are such that the immediate neighborhood of $Pm=1$ is avoided, and  
$Pm<Pm_{c1}<1$ or $Pm>Pm_{c2}>1$.

\begin{figure}
\includegraphics[width=\columnwidth]{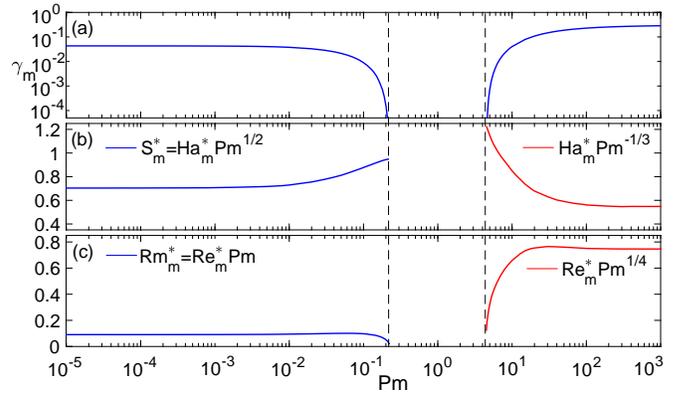}
\caption{Panel (a) shows the growth rate, $\gamma_m$, of type 2 Super-HMRI, 
optimized over a set of parameters $(\alpha, k_z^{\ast}, Ha^{\ast}, Re^{\ast})$, and represented as 
a function of $Pm$, at fixed $Ro=1.5$. Panels (b) and (c) show the corresponding $Ha_m^{\ast}$ 
and $Re_m^{\ast}$, respectively. For both $Pm\ll 1$ and $Pm\gg 1$, $\gamma_m$ tends to
constant values. The re-scaled Hartmann and Reynolds numbers vary as
$Ha_m^{\ast}\propto Pm^{-1/2}$ and $Re_m^{\ast}\propto Pm^{-1}$ for $Pm \ll 1$,
and as $Ha_m^{\ast}\propto Pm^{1/3}$ and $Re_m^{\ast}\propto Pm^{-1/4}$ for $Pm
\gg 1$. Panels (b) and (c) are compensated by these factors to more
clearly highlight these scalings. The dashed lines are at $Pm_{c1}=0.223$
and $Pm_{c2}=4.46$, marking the $Pm=O(1)$ region where no instability exists.}
\end{figure}

\begin{figure}
\includegraphics[width=\columnwidth]{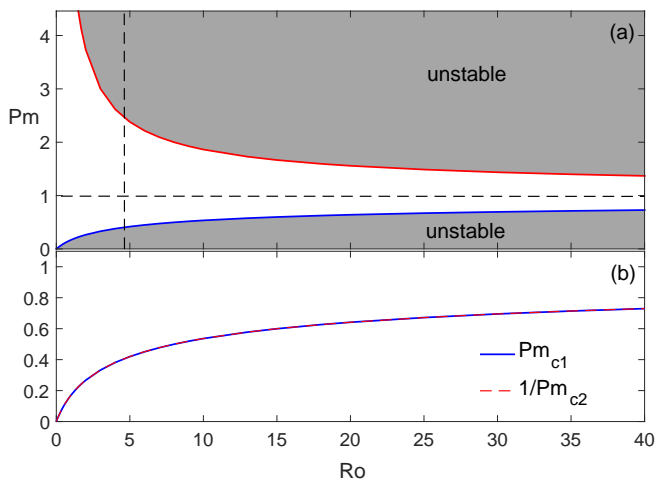}
\caption{Panel (a) shows the lower ($Pm_{c1}$, blue) and upper ($Pm_{c2}$, red) stability 
boundaries of type 2 Super-HMRI vs. $Ro$. The vertical dashed line denotes the Liu limit 
$Ro_{ULL}=4.83$. Panel (b) illustrates that these boundaries are in fact related by 
$Pm_{c1}=1/Pm_{c2}$. Note that this instability exists for both smaller and larger $Ro$ and in fact 
insensitive to the upper Liu limit.}
\end{figure}

Figure 5(a) shows the unstable regions in $(Ro,Pm)$-plane. For all $Ro$, the two bounding curves 
of marginal stability where $\gamma_m=0$ satisfy $Pm_{c1}<1$ and $Pm_{c2}>1$, and are related 
via $Pm_{c1}Pm_{c2}=1$, as seen in Fig.\ 5(b). For high shear ($Ro\rightarrow\infty$), the stability
strip around $Pm=1$ increasingly narrows, so that most $Pm$ values are unstable, 
whereas for $Ro \rightarrow0$, the stable strip widens to include all $Pm$. This is readily 
understood, because shear is the only energy source (just as it is for SMRI, AMRI, and 
HMRI), clearly there can be no instability at all for $Ro=0$, which corresponds to solid-body 
rotation. Note finally how $Pm_{c1}$ and $Pm_{c2}$ stability curves of type 2 Super-HMRI are 
completely unaffected by the Liu limit at $Ro_{ULL}=4.83$, which is otherwise relevant to type 1 
Super-HMRI, and it continues to exist even for $Ro<Ro_{ULL}$.

\begin{figure*}
\includegraphics[width=0.68\columnwidth]{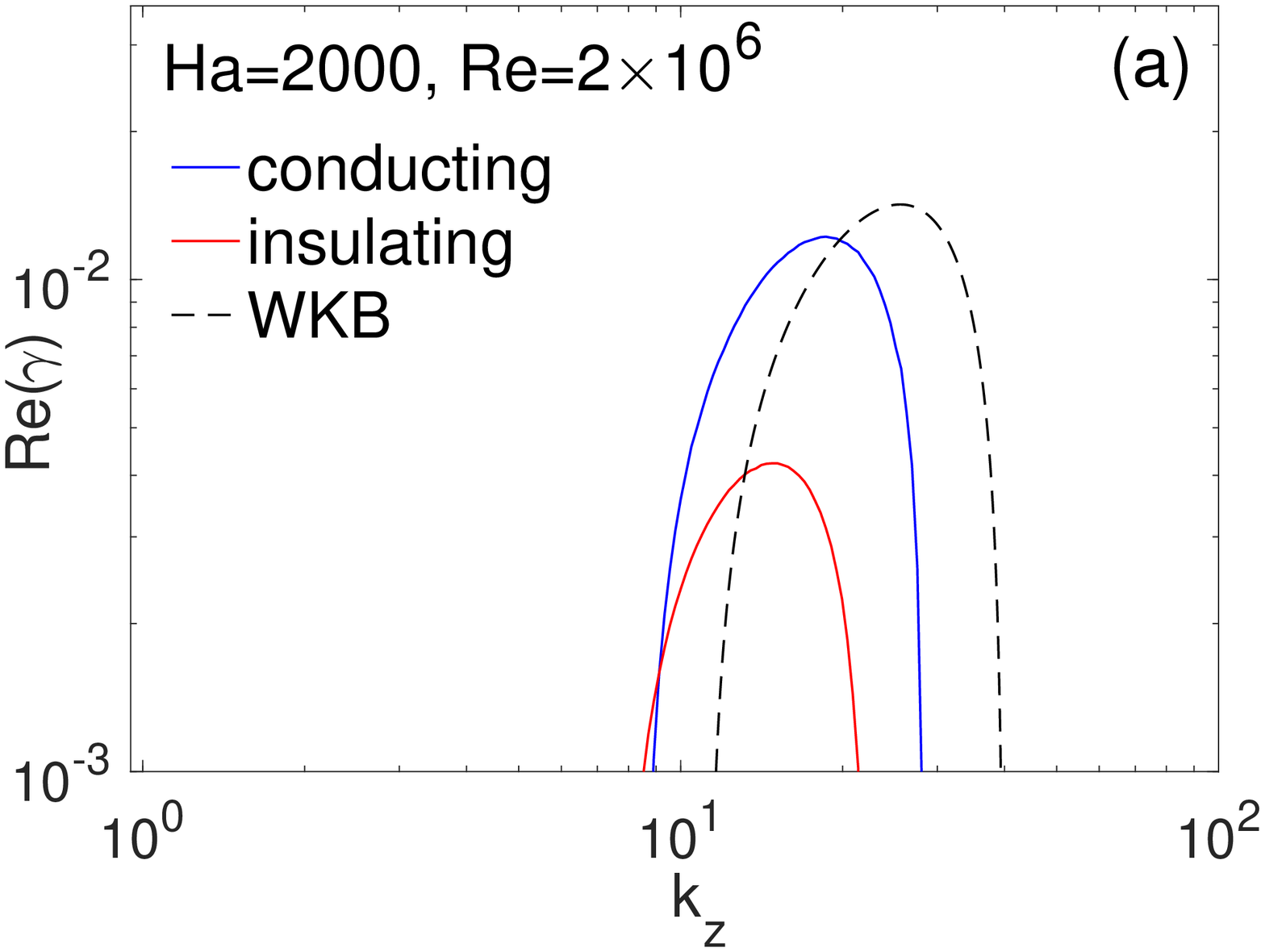}
\includegraphics[width=0.68\columnwidth]{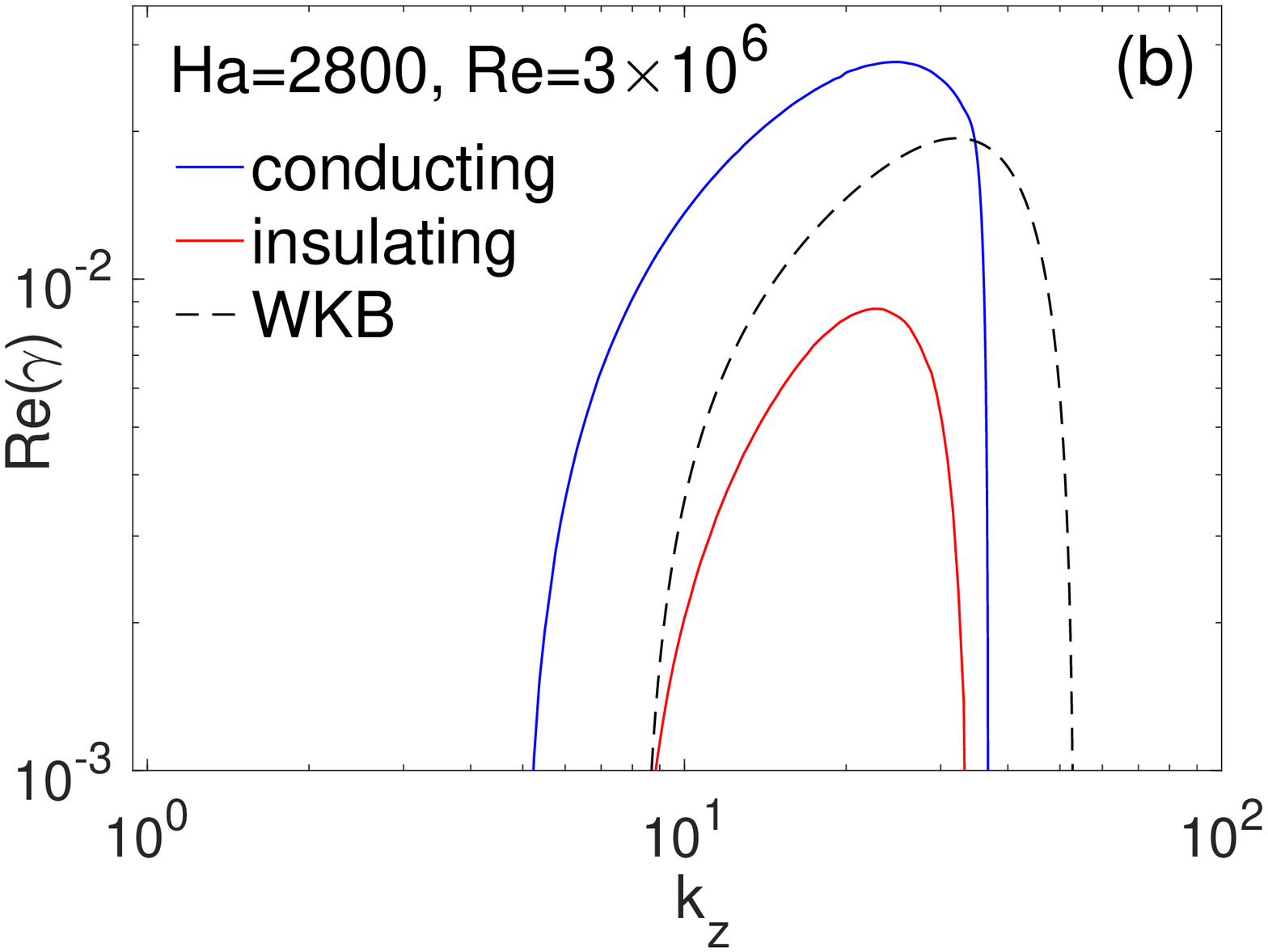}
\includegraphics[width=0.68\columnwidth]{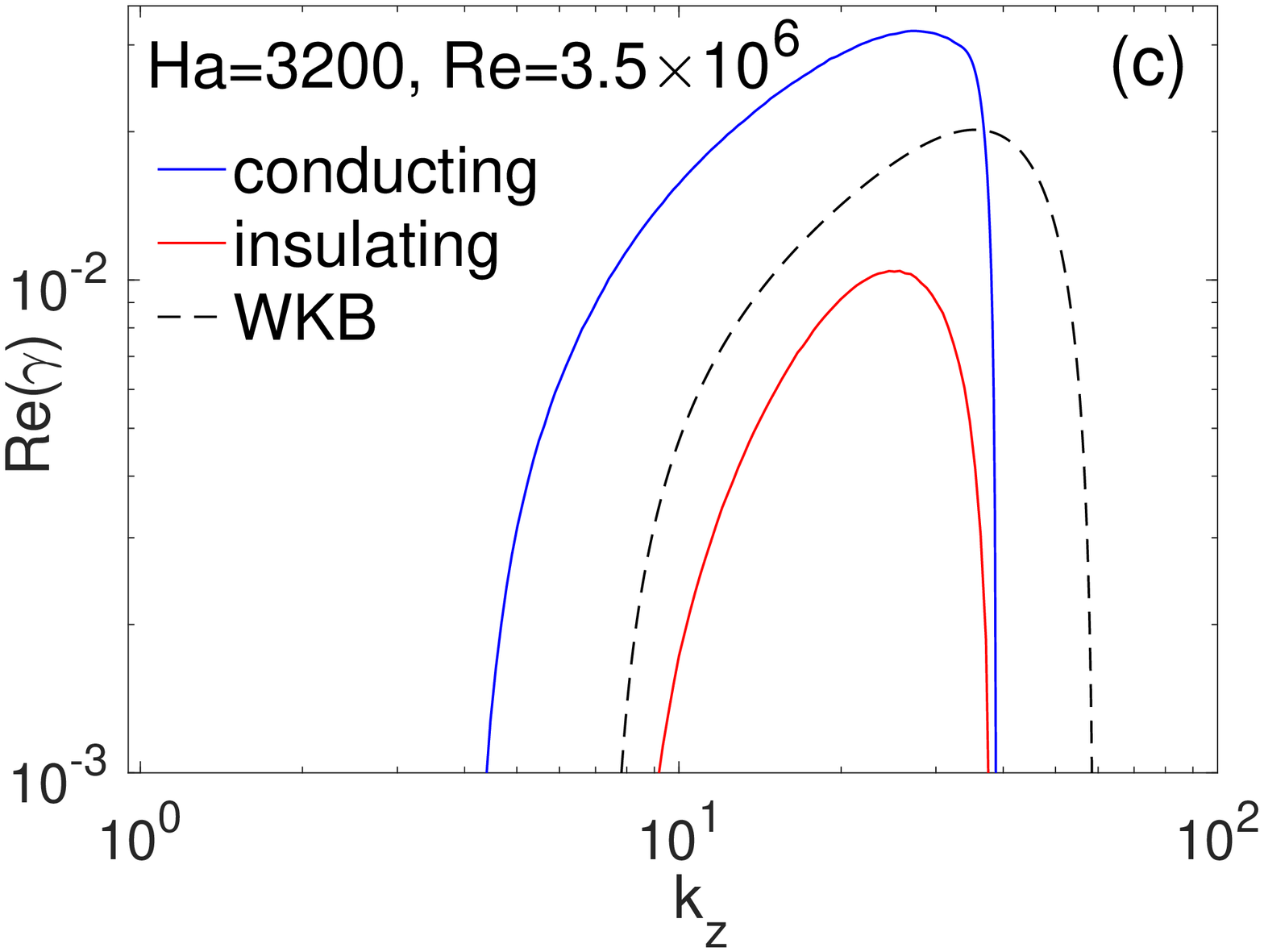}
\includegraphics[width=0.68\columnwidth]{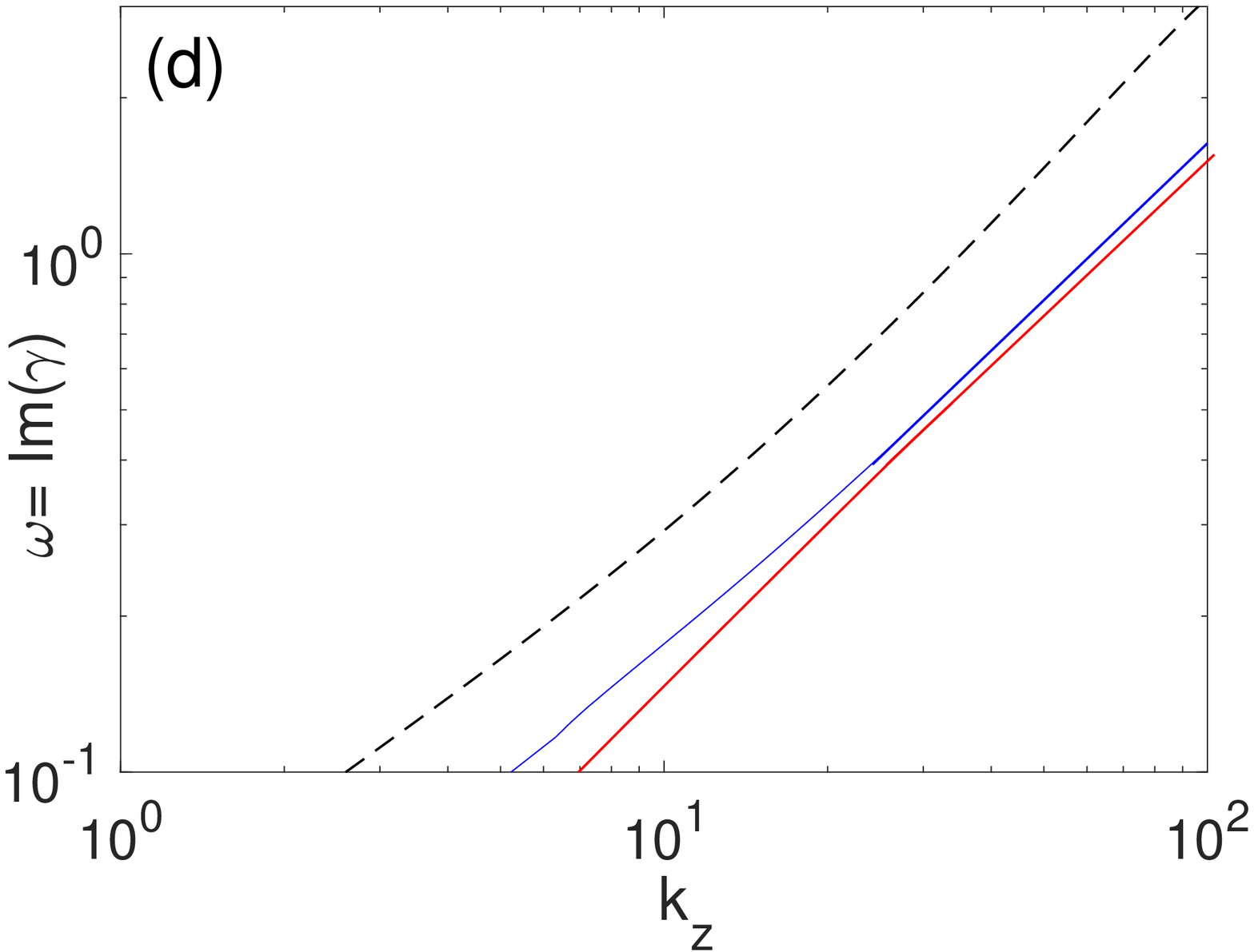}
\includegraphics[width=0.68\columnwidth]{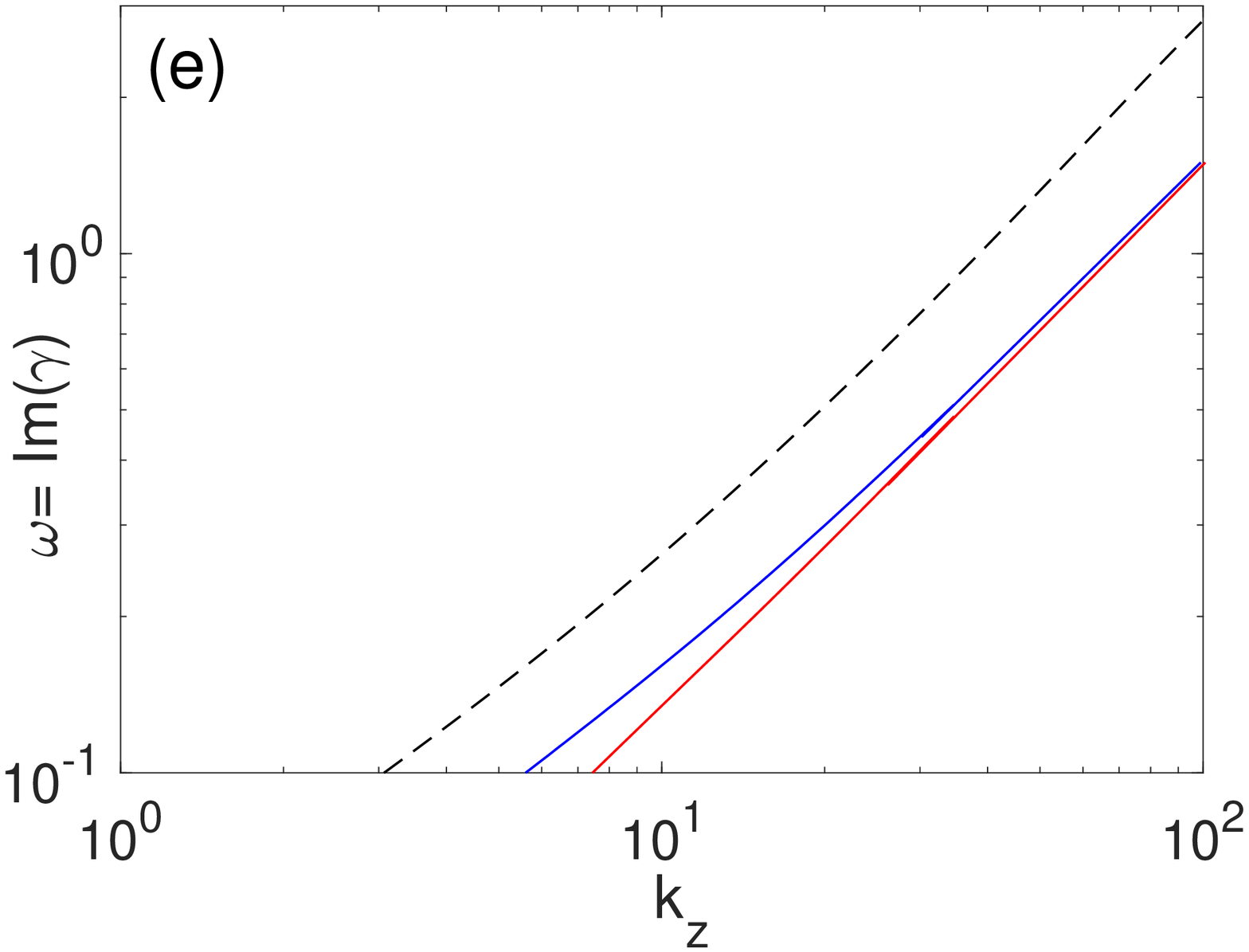}
\includegraphics[width=0.68\columnwidth]{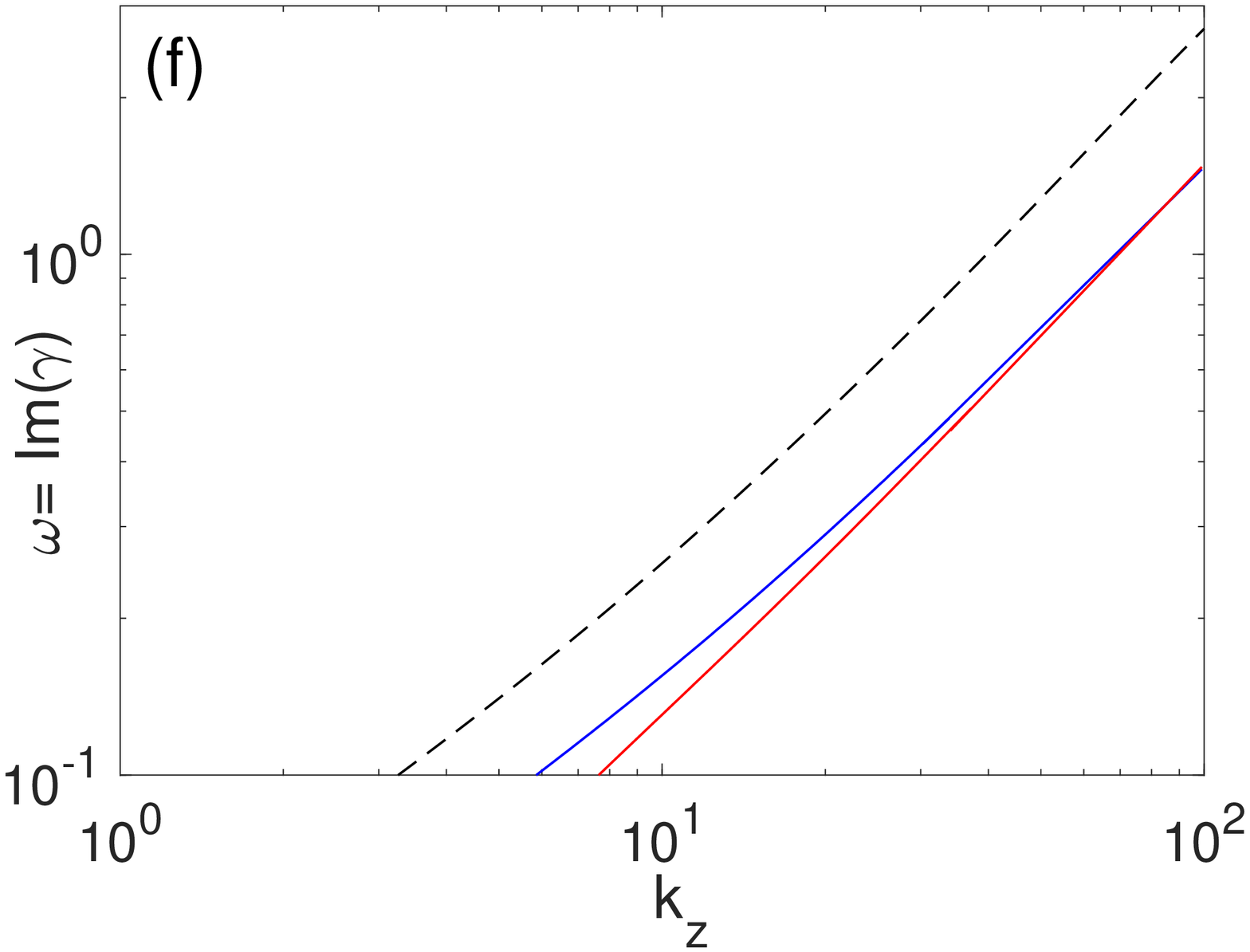}
\caption{Growth rate, $Re(\gamma)$ (panels (a), (b), (c)), and frequency, $\omega={\rm 
Im}(\gamma)$ (panels (d), (e), (f)), vs. $k_z$ in the global case with the narrow gap $\hat{\eta}=0.85$ and 
power-law angular velocity profile with constant $Ro=3.5 < Ro_{ULL}$ in the presence of 
conducting (blue) and insulating (red) boundary conditions imposed on the cylinders. In all panels, 
$\beta=100$, $Pm=10^{-3}$, while $(Ha, Re)=(2000,~2\times10^6)$ in panels (a) and (d),  $(Ha, 
Re)=(2800,~3\times10^6)$ in panels (b) and (e), and $(Ha, Re)=(3200,~3.5\times10^6)$ in panels 
(c) and (f). Black dashed curve in each panel is accordingly the growth rate or frequency resulting 
from the local WKB dispersion relation (Eq.\ 4) for the same values of the parameters corresponding to that 
panel and a fixed $k_{r0}=\pi/\delta$, where $\delta=1-\hat{\eta}$ is the gap width in units of $r_o$. 
It is seen that in the local and global cases, the shape of the dispersion curves is qualitatively same, with comparable
growth rates, frequencies and corresponding wavenumbers, however, the boundary conditions tend to shift the growth rates towards 
lower axial wavenumbers, while the frequencies towards larger wavenumbers relative to those in the 
local case. Also, insulating boundaries lower the growth rate about three times. As for the 
frequencies, they are close to each other for both these boundaries and smaller than those in the 
local case at a given $k_z$.}
\end{figure*}

\begin{figure*}
\includegraphics[width=0.68\columnwidth]{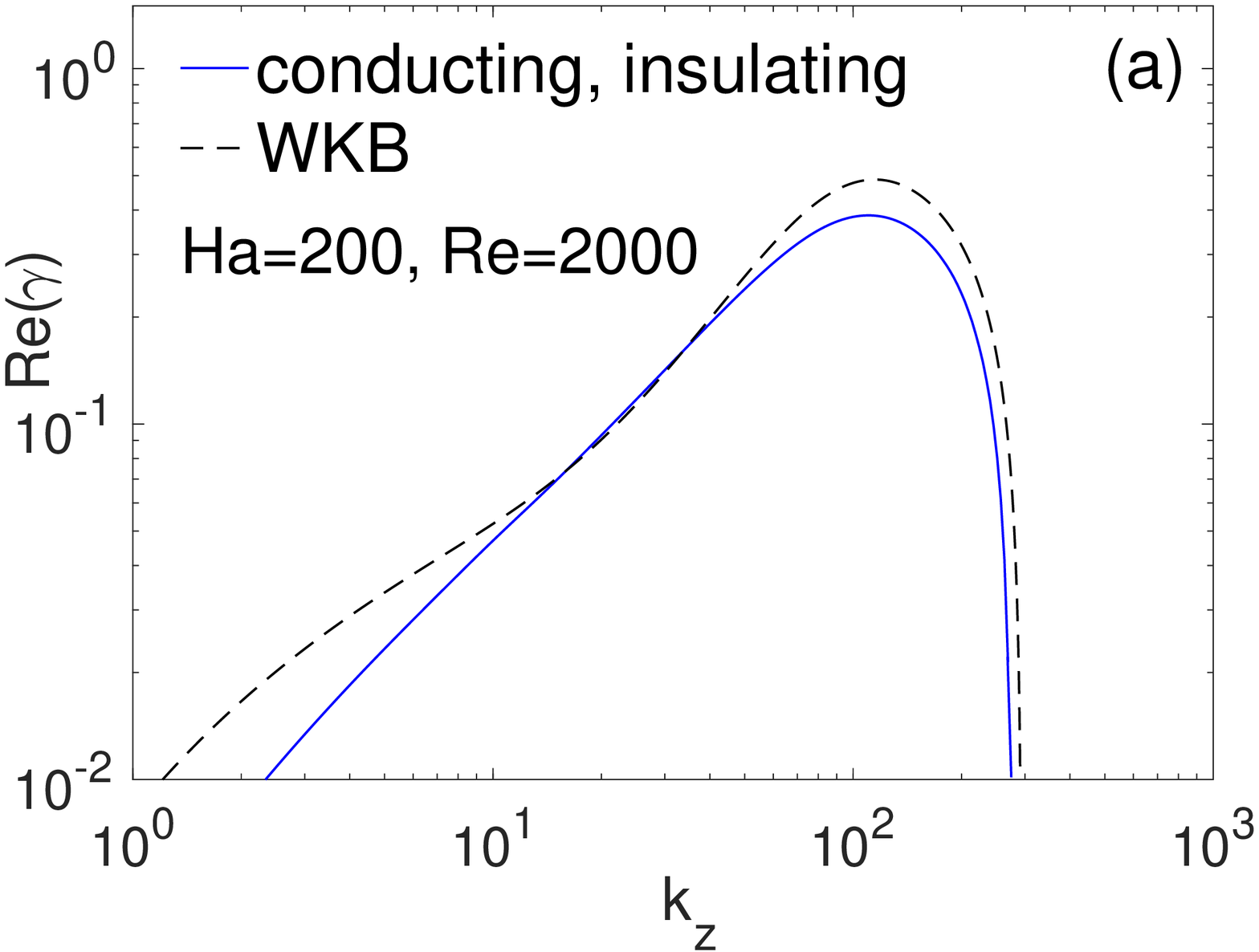}
\includegraphics[width=0.68\columnwidth]{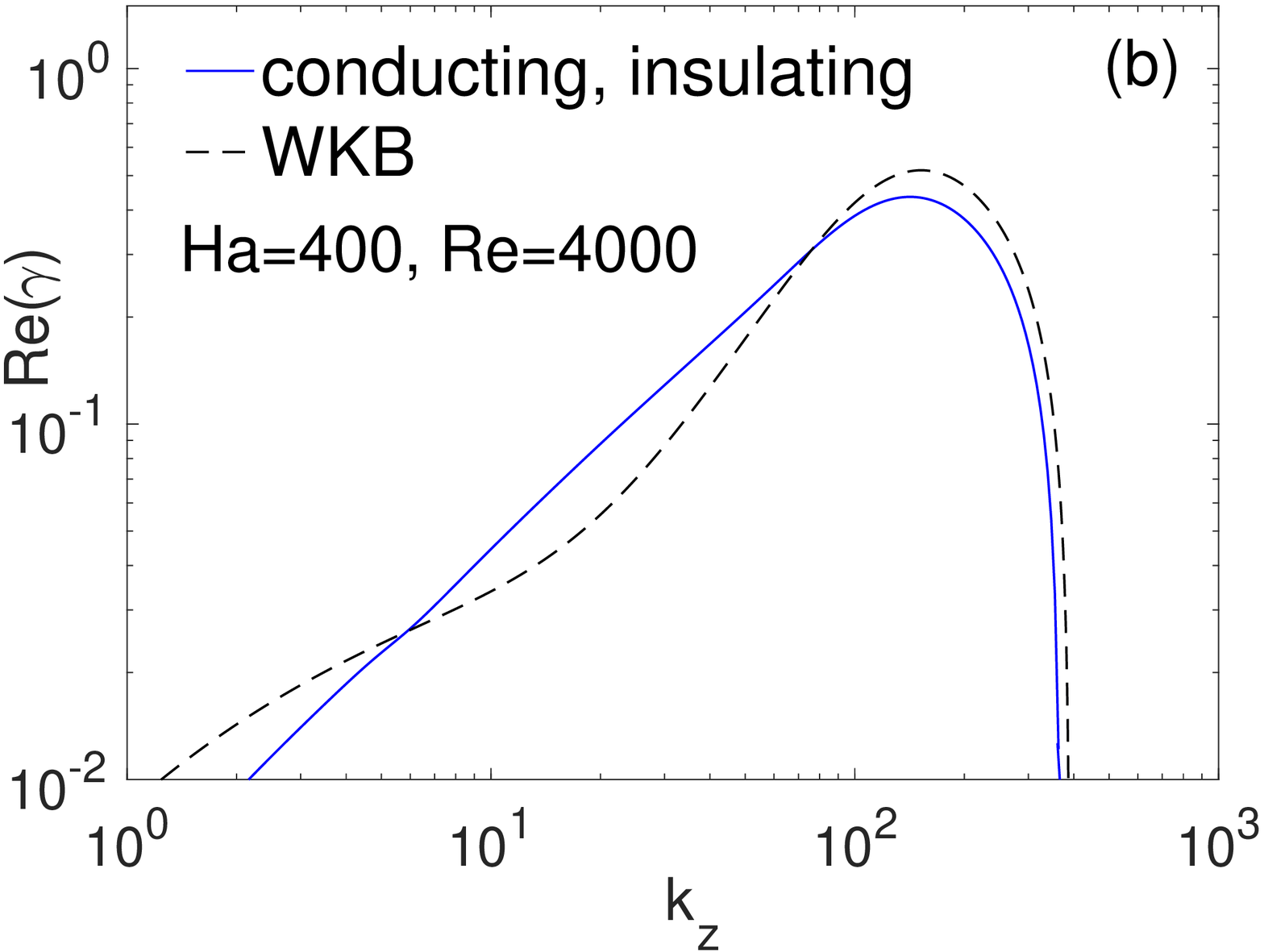}
\includegraphics[width=0.68\columnwidth]{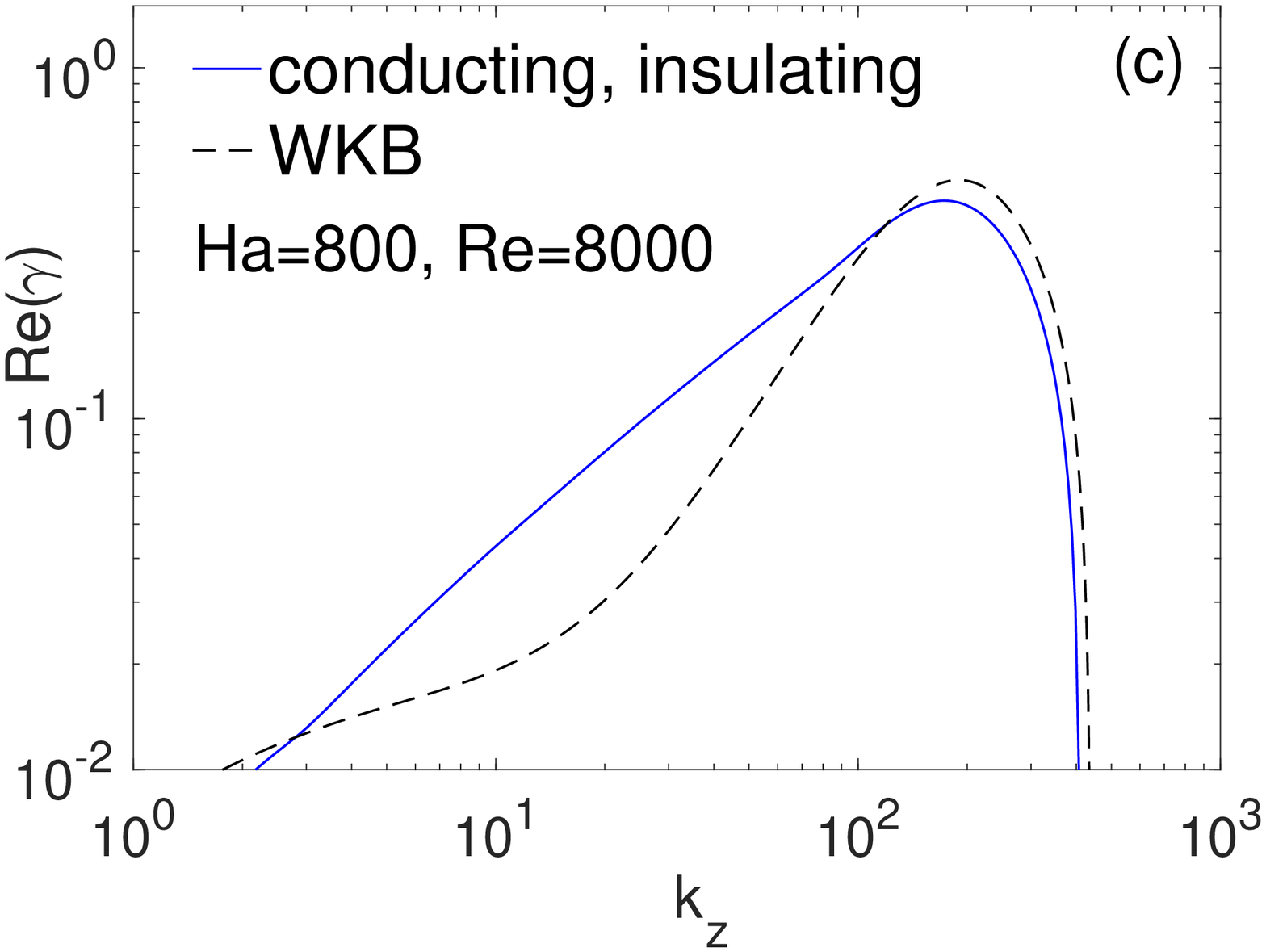}
\includegraphics[width=0.68\columnwidth]{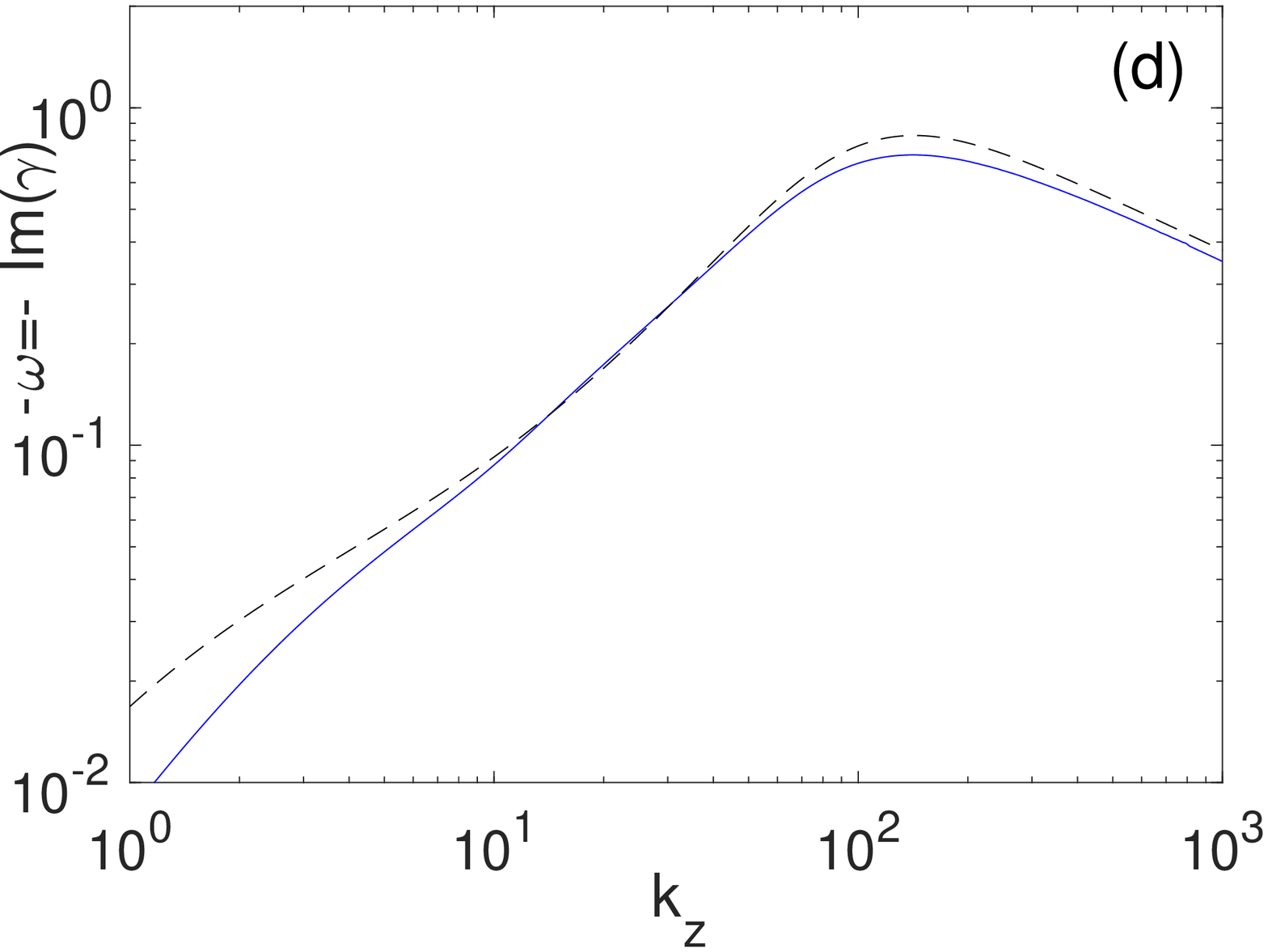}
\includegraphics[width=0.68\columnwidth]{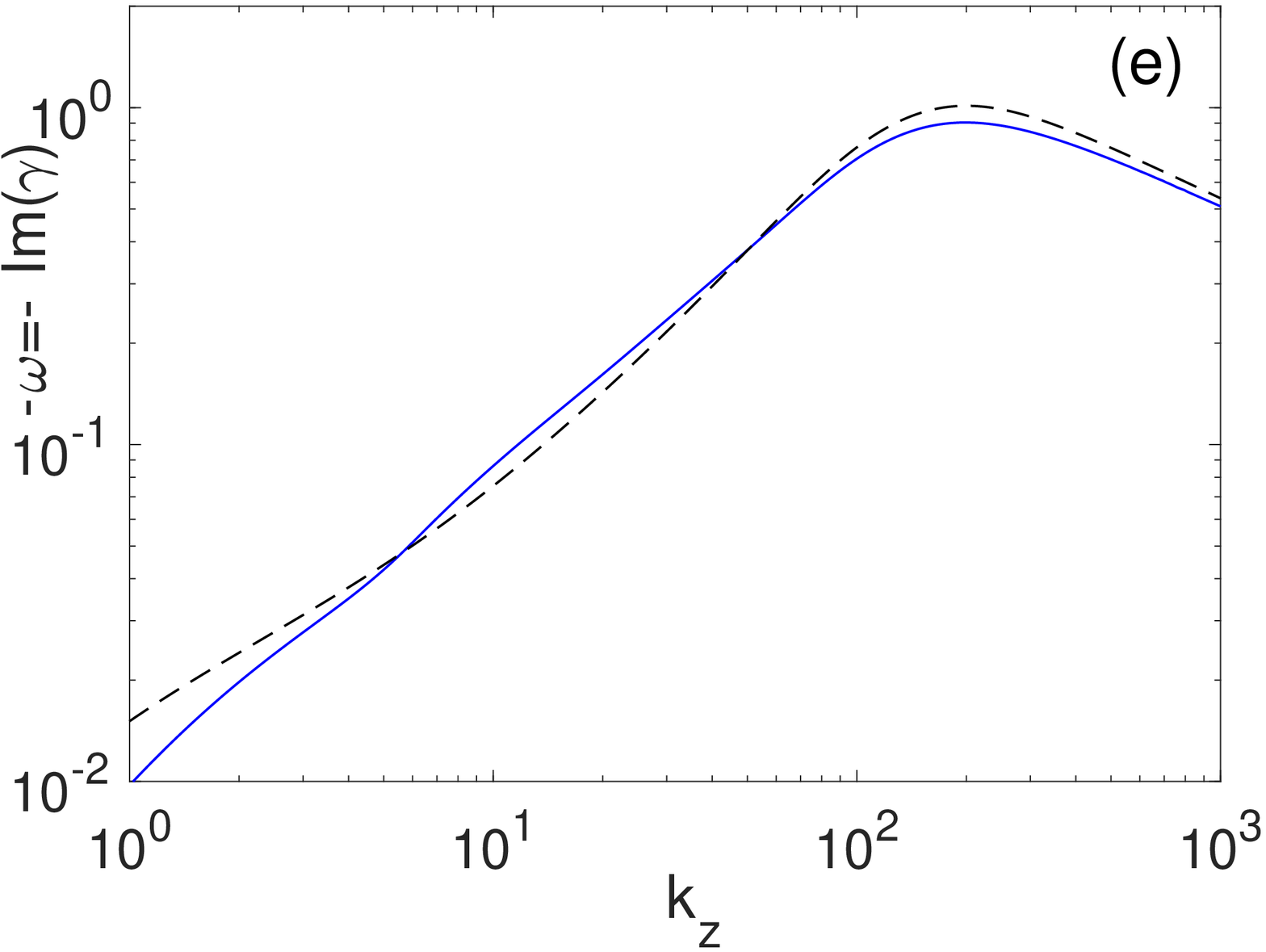}
\includegraphics[width=0.68\columnwidth]{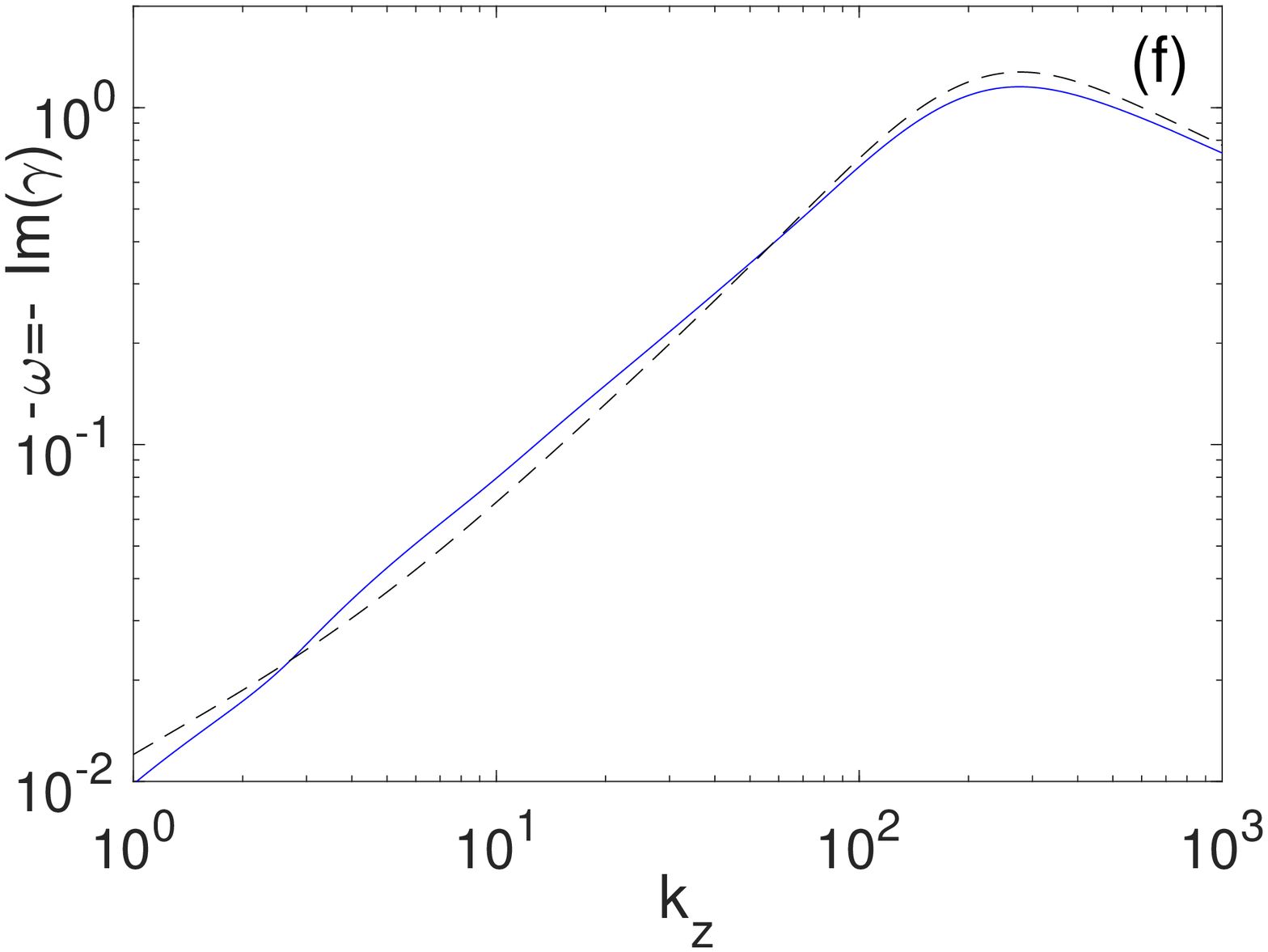}
\caption{Same as in Fig. 6, but at large $Pm=100$ with $(Ha, Re)=(200,~2000)$ in panels (a) and 
(d), $(Ha, Re)=(400,~4000)$ in panels (b) and (e), and $(Ha, Re)=(800,~8000)$ in panels (c) and 
(f). Note that in this case the growth rates and frequencies (plotted is $-\omega=-{\rm 
Im}(\gamma)$) are nearly indistinguishable for conducting and insulating boundary conditions, in 
contrast to the small-$Pm$ regime depicted in Fig. 6. Black dashed curve in each panel is again 
accordingly the growth rate or frequency resulting from the WKB dispersion relation (Eq.\ 4) with 
$k_{r0}=\pi/\delta$ and the same values of the parameters corresponding to that panel. For the 
growth rate, the difference between the local and global dispersion curves increases with 
increasing Hartmann and Reynolds numbers mainly at smaller $k_z\lesssim 10$ (panel (c)), 
whereas for the frequencies these curves are quite close to each other. Overall, the agreement 
between the WKB and 1D analysis at these large $Pm=100$ is much better 
than that at small $Pm=10^{-3}$ in Fig. 6. }
\end{figure*}

\begin{figure}
\includegraphics[width=\columnwidth]{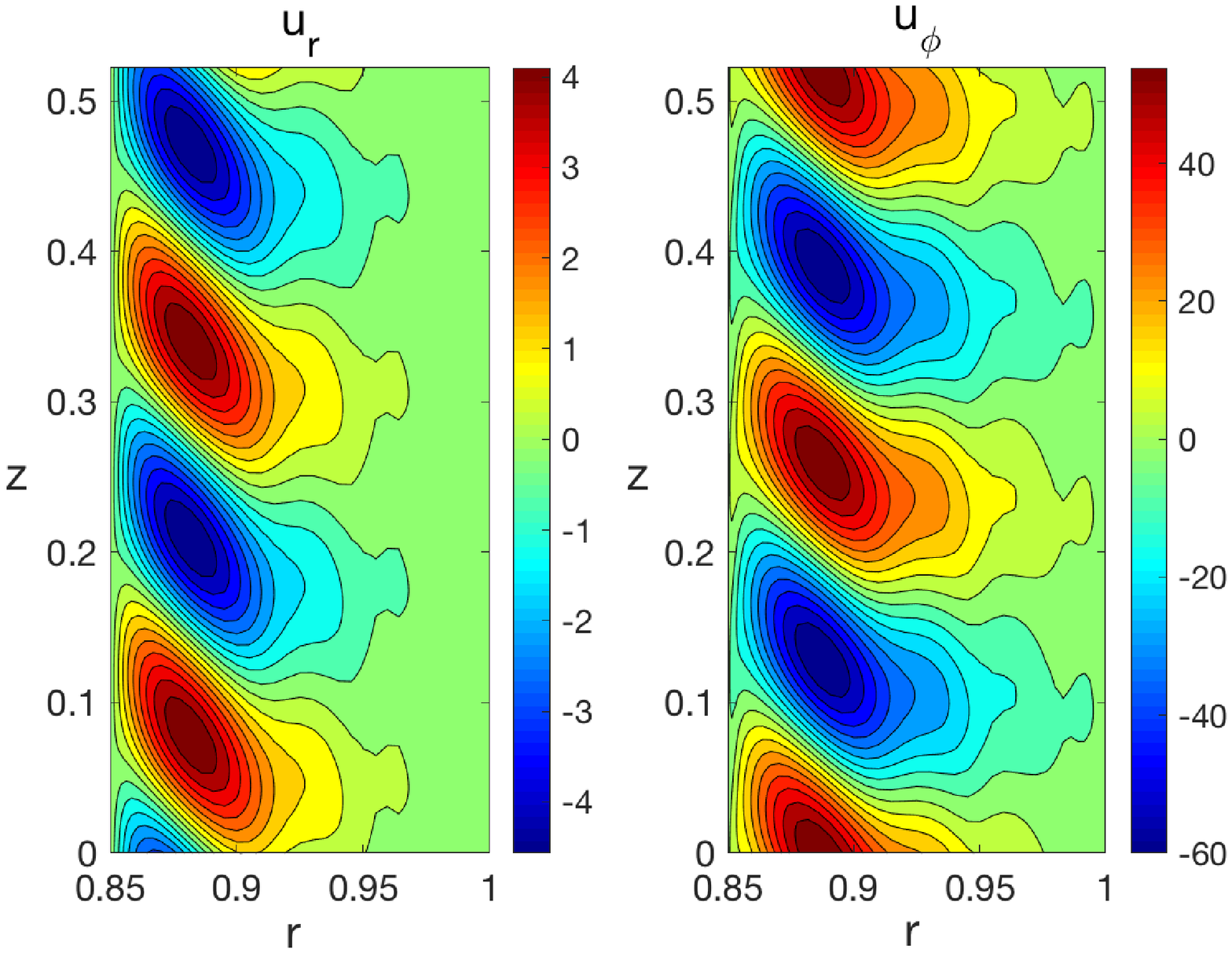}
\includegraphics[width=\columnwidth]{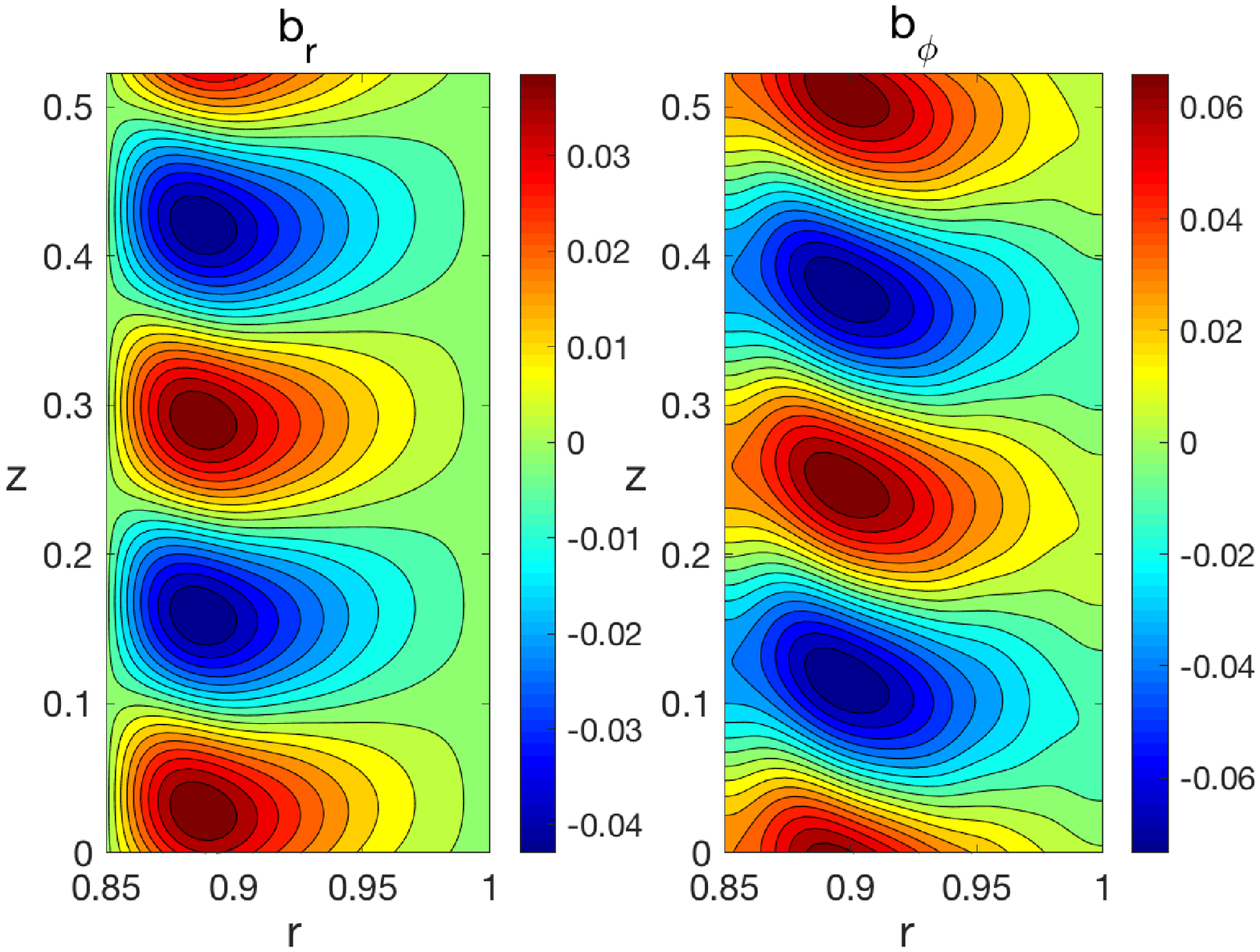}
\caption{Eigenfunctions of the radial and azimuthal velocity and magnetic field for type 2 
Super-HMRI in the global case as a function of $(r, z)$ at small $Pm=10^{-3}$, the narrow gap 
$\hat{\eta}=0.85$ and conducting boundary conditions on the cylinders. 
The parameters $(\beta, Ro, Ha, Re)$ are the same as those in Fig.\ 6(b) and the 
axial wavenumber is chosen to be $k_{zm}=24$, at which the growth rate for the given values of 
these parameters reaches a maximum (peak on the blue curve in that panel).}
\end{figure}

\begin{figure}
\includegraphics[width=\columnwidth]{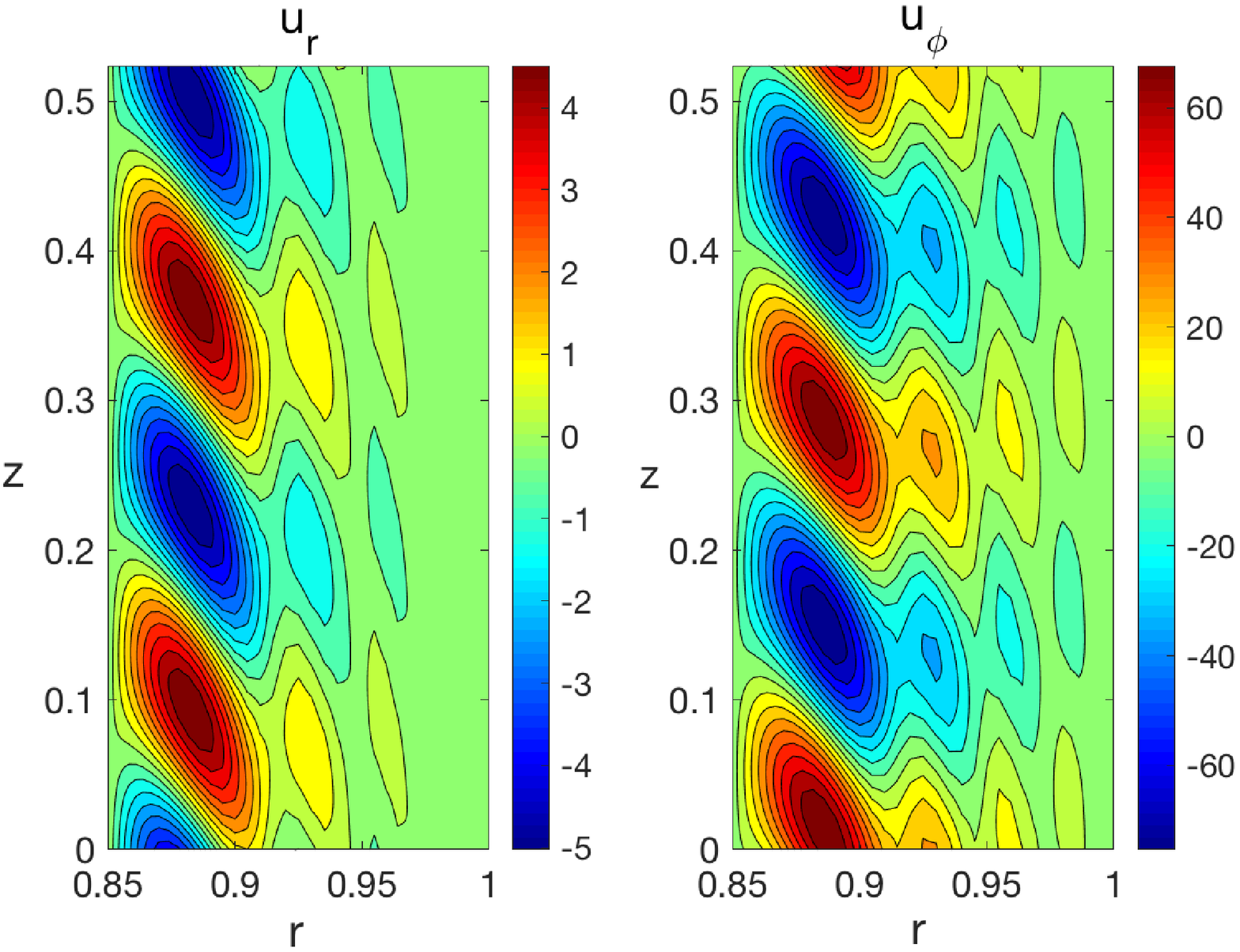}
\includegraphics[width=\columnwidth]{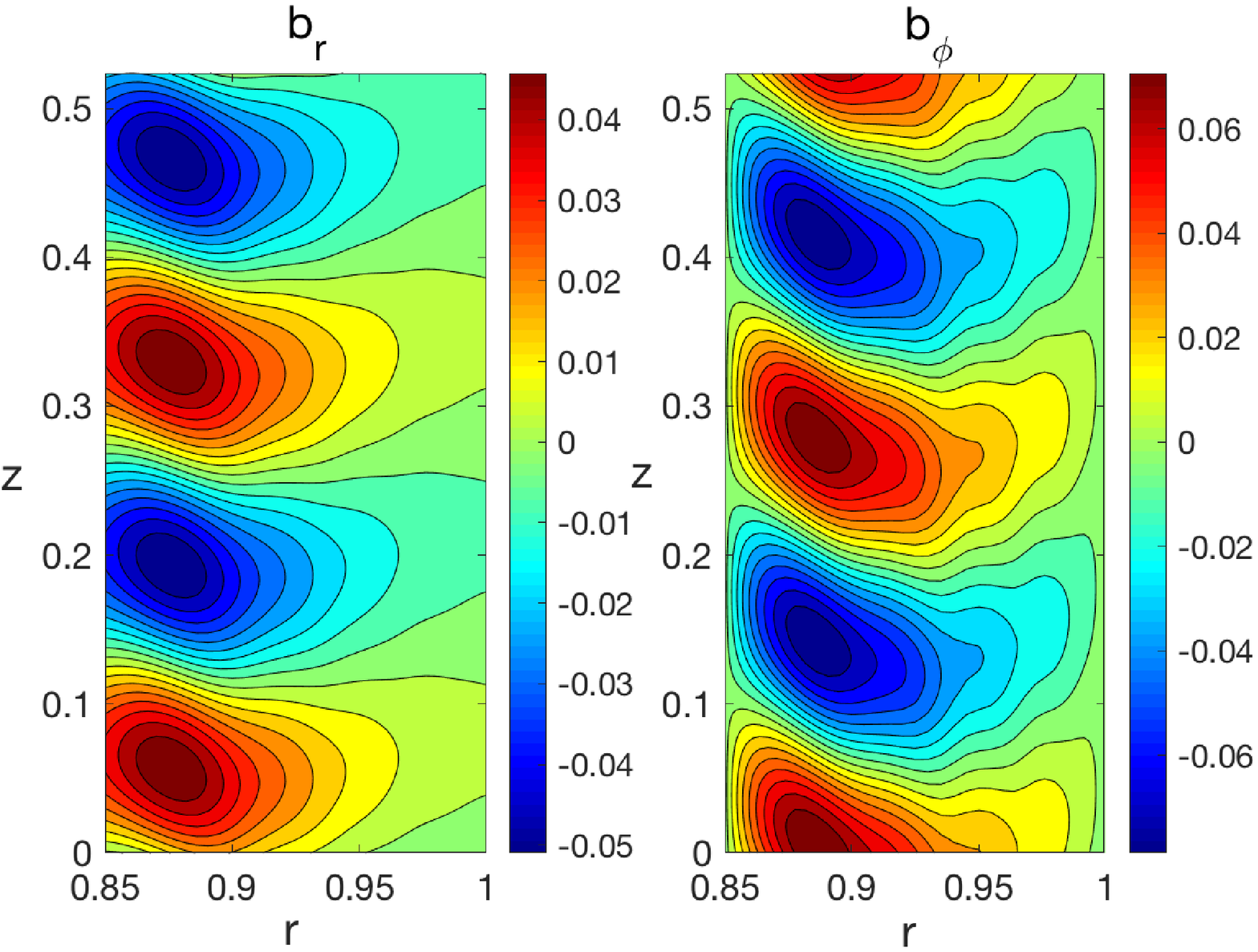}
\caption{Same as in Fig.\ 8, but for insulating boundary conditions on the cylinders. The main 
parameters are again the same as in Fig.\ 6(b), but now the axial wavenumber is chosen to be 
$k_{zm}=23$, corresponding to the largest growth rate for the given values of these parameters 
(peak on the red curve in that panel) and the boundary conditions.}
\end{figure}

\begin{figure}
\includegraphics[width=\columnwidth]{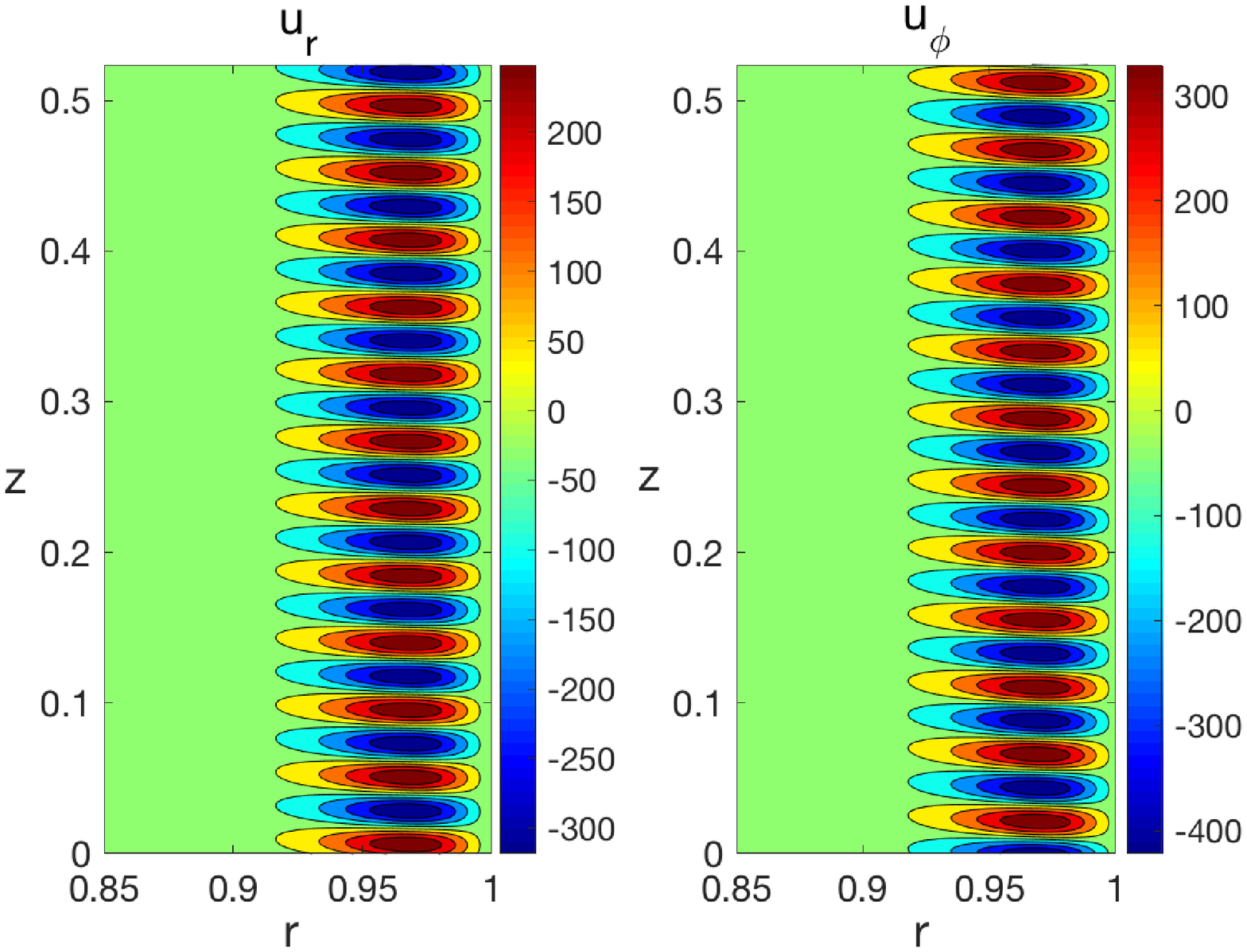}
\includegraphics[width=\columnwidth]{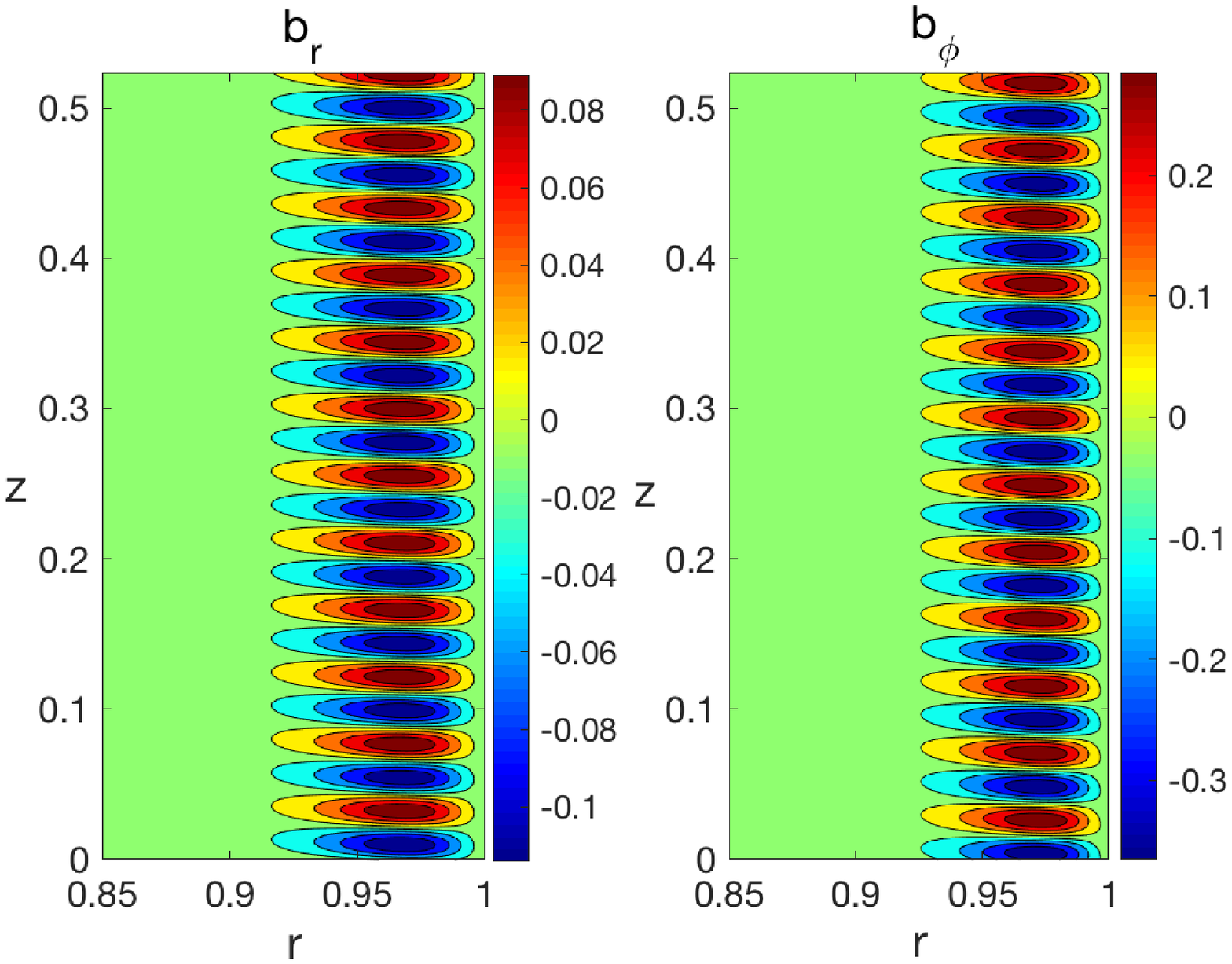}
\caption{Same as in Fig.\ 8, but for large $Pm=100$ and conducting boundary conditions. The 
parameters $(\beta, Ro, Ha, Re)$ are the same as in Fig.\ 7(b) and the axial wavenumber is 
$k_{zm}=141$, corresponding to the maximum growth rate (peak on the blue curve in that panel) for the given 
values of these parameters. For the sake of comparison of the eigenfunction structures and 
lengthscales with those at small $Pm$, the $z$-axis has the same range as in Figs.\ 8 and 9. 
At large $Pm$, the eigenfunctions in the presence of insulating boundaries are in fact identical to
the ones for the conducting boundaries shown here at the same values of the main parameters 
and hence we do not plot those eigenfunctions.}
\end{figure}

Having described the properties of type 2 Super-HMRI using the WKB analysis, 
we should now address the question of validity of this approach. A basic condition for the 
local WKB approximation to hold is that the radial wavelength, $\lambda_r$, of the 
perturbations must be much smaller than the characteristic radial size, $r_o$, of the system over 
which the equilibrium quantities vary, i.e., $\lambda_r\ll r_o$. In terms of the radial wavenumber 
$k_r=2\pi/\lambda_r$, this condition becomes $r_ok_r\gg 1$, or in non-dimensional units used here 
$k_r\gg 1$. From the above analysis (see Figs. 1(a) and 2(a)), it appears that the re-scaled 
wavenumbers typical of this instability are relatively small, $k_r^{\ast}\sim k_z^{\ast} \lesssim 
0.1-1$. Recalling that $k_r=\beta k^{\ast}_r$, it follows that WKB approach is strictly applicable 
only at large $\beta \gg 1/k_r^{\ast} \gtrsim 10$, i.e., at dominant background azimuthal 
magnetic field. However, we will see in the global linear analysis below that this mode of instability 
is in fact not restricted only to large $\beta$ and can even exist at smaller $\beta \sim 1$, but in 
this case the WKB approach is questionable and should be applied with caution.

\section{Global analysis}

After the radially local WKB analysis, we now investigate type 2 Super-HMRI in the global case. 
The problem reduces to a 1D (along radius) linear eigenvalue problem in a viscous 
and resistive rotational flow between two coaxial cylinders threaded by helical magnetic field, as 
outlined in \cite{Hollerbach_Ruediger2005}. Hence, the main equations are the same as those in 
that paper and are obtained by linearizing Eqs.\ (1)-(3) about the above equilibrium, with the 
only difference being that now the imposed rotation profile $\Omega(r)$ increases with radius 
corresponding  to outer cylinder rotating faster than the inner one (super-rotation). The boundary 
conditions are no-slip for the velocity and either perfectly conducting or insulating for the magnetic field. 
As in \cite{Hollerbach_Ruediger2005} (see also \cite{Priede_etal2007, Priede2011}), the radial 
structure of the quantities are expanded in Chebyshev polynomials, typically up to $N=30-40$. 
The governing equations and boundary conditions then reduce to a large ($4N\times 4N$) matrix 
eigenvalue problem, with the real parts of eigenvalues being growth rates of the eigenmodes and 
imaginary parts their frequencies (see also Ref. \cite{Hollerbach2007} for the details of the 
numerical scheme). In this paper, the global 1D analysis serves two main purposes. First, to 
compare with the results of the above WKB analysis in the regime where the latter holds, i.e., at 
high $\beta\gg 10$ and a small gap width between the cylinders, $\delta\equiv r_o-r_i \ll r_o$, 
when the equilibrium quantities do not change much with radius across the gap. Such an analysis 
will allow us to ascertain the existence of type 2 Super-HMRI also in the global setup and to 
characterize influence of the boundary conditions (conducting, insulating) for the magnetic field imposed on the 
cylinders. Second, with a view to detecting this instability in the upcoming liquid sodium TC 
experiments at the DRESDYN facility \cite{Stefani_etal2019}, we look for it at a larger gap width and 
smaller $\beta \sim 1$ -- a regime where WKB approximation becomes questionable, but, on the 
other hand, is more relevant to experimental conditions. We start our global linear stability analysis 
with the first case. 

\subsection{Narrow-gap case}

One of the main difficulties when comparing local and global analysis of HMRI and AMRI in a 
magnetic TC flow is that the local Rossby number, which defines these instabilities and to which 
they are therefore sensitive \cite{Liu_etal2006,Kirillov_Stefani2010,Kirillov_etal2014}, varies with 
radius, even in the narrow gap case. On the other hand, it is assumed to be radially constant in the 
WKB treatment, as other equilibrium variables are. So, there is some uncertainty in matching Rossby 
numbers in the local and global analysis \cite{Stefani_Kirillov2015}. To circumvent this problem and 
facilitate comparison with the WKB analysis, following \cite{Ruediger_Hollerbach2007}, we impose 
a power-law radial profile of angular velocity with a constant positive Rossby number,
\begin{equation}
\Omega(r)=\Omega_o\left(\frac{r}{r_o}\right)^{2Ro},
\end{equation}
(outer cylinder's $\Omega_o=1$ in our units) as a background flow between the cylinders instead 
of the usual TC flow profile. The gap is assumed to be narrow 
with the ratio of inner to outer cylinders' radii $\hat{\eta}=r_i/r_o=0.85$. The Rossby number is 
chosen smaller than the upper Liu limit, $Ro=3.5 < Ro_{ULL}$, thereby excluding type 1 
Super-HMRI in the global case too and allowing us to focus only on the new type 2 Super-HMRI 
branch (the results are similar also at smaller $Ro$). 

From the WKB analysis above, we have seen  that for a given value of the growth rate, 
the associated wavenumbers are proportional to $\beta$ helicity parameter. As a result, in order to confine type 2 Super-HMRI within the gap extent in the 
global case and at the same time remain in the domain of validity of the WKB approximation, its 
radial wavelength should be at least comparable to or smaller than the gap width and, 
consequently, much smaller than the outer cylinder radius $r_o$. In terms of the radial wavenumber 
this condition becomes, $k_r \gtrsim \delta^{-1}$, which in our non-dimensional units turns out to 
be large $k_r \gtrsim r_o\delta^{-1}=(1-\hat{\eta})^{-1}\gg 1$, because the gap is narrow 
$\hat{\eta}\approx 1$ (similarly increases with $\beta$ also the vertical wavenumber $k_z$). In 
terms of the re-scaled wavenumbers used in the WKB analysis, we get $\beta 
k_r^{\ast} \gtrsim (1-\hat{\eta})^{-1}=6.67$, with which we can estimate from Figs. 1(a) and 2(a)
that in order to get such high wavenumbers, we arrive on the same constraint that $\beta$ should 
be large, $\beta  \gtrsim 100$ (for the adopted values of $Ro$ and $Pm$). It also follows 
from Figs.\ 1(c) and 2(c) that due to the above scaling with $\beta$, the Hartmann and Reynolds numbers, for which 
the instability grows most, are accordingly also fairly high: $Ha\sim 10^4$, $Re \sim 10^9$ for 
small $Pm=10^{-6}$ and $Ha\sim 100$, $Re\sim 10^3$ for large $Pm=100$. 
In order to avoid the numerical difficulties in our 1D eigenvalue code when both 
axial wavenumber and Reynolds number are such high ($k_z \gtrsim 10$ and $Re\sim 10^9$), we take 
higher $Pm=10^{-3}$ in these global calculations at the narrow gap instead of 
$Pm\sim 10^{-6}$ used in Fig.\ 1. This, however, does not change the regime of the instability, 
because according to Fig.\ 4, we are still in the same asymptotic regime of low $Pm \ll 1$, when 
the growth rate is determined by $S$ and $Rm$. As a result, the adopted $Pm=10^{-3}$ yields 
lower values of $Ha\sim 10^3$ and $Re \sim 10^6$ for the same growth rate, which is 
computationally not as challenging. Such numerical problems do not arise for $Pm=100$, where 
$Re$ is three orders of magnitude lower, so we keep this value of the magnetic Prandtl number in 
the global analysis for the case when it is high and the gap is narrow. 

Figures 6 and 7 present the results of the global calculations in the narrow gap case. Plotted here 
is the growth rate and frequency of type 2 Super-HMRI as a function of $k_z$ at conducting and 
insulating boundary conditions for, respectively, small $Pm=10^{-3}$ and large $Pm=100$ and 
various pairs of $(Ha, Re)$. For comparison, we also show the solutions of the WKB dispersion 
relation (Eq.\ 4) at a fixed radial wavenumber $k_{r0}=\pi(1-\hat{\eta})^{-1}$ in these figures. This 
choice of the radial wavenumber when comparing local results with 1D global ones is dictated by 
the fact that global eigenfunctions (see below) usually extend almost over the whole domain 
between the cylinders and do not change sign in the radial direction within this domain (see also 
Refs. \cite{Ji_etal2001, Stefani_Kirillov2015}) \footnote{Actually the selected value of $k_{r0}$ 
when comparing local WKB and global dispersion curves is still somewhat arbitrary, since it is not 
generally possible to get an exact matching between the characteristic radial length of the global 
eigenfunctions and radial wavenumber of WKB solutions/harmonics.}. It is seen that in all these 
cases, both the frequency and growth rate of the instability exhibit generally a qualitatively similar 
dependence on the axial wavenumber as in the local analysis, with comparable maximum growth 
rates and corresponding $k_{zm}$, however, there is a noticeable quantitative difference between 
the small and high-$Pm$ regimes -- the influence of the boundary conditions on the frequency and 
growth rate and hence deviations from the WKB results are larger at small $Pm$. The boundary 
conditions cause the dispersion curves for the growth rate to somewhat shift towards lower $k_z$, 
while for the frequencies towards larger $k_z$ relative to those in the local case. Besides, 
depending on $Ha$ and $Re$, conducting boundaries can lead either to increase or decrease of 
the growth rate compared to its WKB value, whereas insulating boundaries always reduce the 
growth rate about three times compared to those for the conducting ones for fixed Hartmann and 
Reynolds numbers. As for the frequencies, they are quite close to each other for both boundary 
conditions and, at a given $k_z$, smaller than those in the local case. By contrast, for large $Pm$, the growth rate and 
frequency are essentially the same (dispersion curves are indistinguishable) for conducting and 
insulating boundaries (for this reason, Fig.\ 7 shows only the blue curves that represent both the 
boundary conditions) and quite close to their respective values from the WKB analysis, although 
with increasing $Ha$ and $Re$ the deviation from the local analysis becomes more and more 
noticeable, especially for the growth rate at smaller $k_z$ (Fig.\ 7(c)). These behavior with $Pm$ is 
also consistent with previous studies on Super-AMRI at positive shear, indicating generally 
higher critical onset values of Hartmann and Reynolds numbers and, for given values of these numbers, 
smaller growth rates for insulating boundaries than those for conducting ones at smaller 
$Pm$, but almost identical critical values of $Ha$ and $Re$ and growth rates for both these boundary 
conditions at large $Pm$ \cite{Ruediger_etal2018b,Ruediger_etal2018c}. 

To better understand and interpret this behavior of type 2 Super-HMRI in the global case, we 
computed the associated spatial eigenfunctions. The structure of these eigenfunctions for the 
radial and azimuthal components of the velocity and magnetic field in $(r,z)$-plane at small and 
large $Pm$, both for the conducting and insulating boundary conditions, are shown in Figs.\ 8-10. The 
values of axial wavenumbers in each figure are chosen such that to get the largest growth rate for 
given values of the remaining parameters in Figs.\ 6(b) and 7(b), i.e., $k_{zm}=24$ in the presence of
conducting and $k_{zm}=23$ of insulating boundaries at small 
$Pm=10^{-3}$, while at large $Pm=100$, we take $k_{zm}=141$, which is the same for both the 
boundary conditions. First of all, we note that despite differences in structure, these eigenfunctions 
appear quite similar to those of the more familiar HMRI and AMRI in a TC setup with negative 
shear (e.g., \cite{Hollerbach_Ruediger2005, Hollerbach_etal2010, Guseva_etal2015, 
Mamatsashvili_etal2018,Ruediger_etal2018a}), extending over most part of the radial extent of the flow. 
They do not display any strong concentration in the vicinity of only the inner or 
outer radial boundary, which would otherwise mean that the instability mode is induced due to 
those boundaries. Thus, although the radial boundaries, either insulating or conducting, affect the 
growth rate, especially at small $Pm$ (Fig.\ 6), they are not the main cause/driver of type 2 
Super-HMRI. It is the combination of shear, helical magnetic field and dissipation that gives rise to 
this instability. In this regard, it is already known that the type of radial boundary conditions for the 
magnetic field also plays an important role in the dynamics of more extensively studied ``relative'' 
-- axisymmetric HMRI and non-axisymmetric AMRI at negative shear \cite{Priede_etal2007, 
Ruediger_Hollerbach2007, Ruediger_etal2018c}, whose basic physics can be understood 
within the local WKB analysis \cite{Liu_etal2006,Kirillov_Stefani2010,Kirillov_Stefani2012,Priede2011,Kirillov_etal2014}. 
 
It is seen in these figures that the eigenfunctions markedly differ in structure depending on the 
magnetic Prandtl number and boundary conditions. At $Pm=10^{-3}$, they appear to have 
comparable axial and radial lengthscales, which are of the same order as the radial size of the gap, 
with the radial scale of the velocities being somewhat smaller than that of the magnetic field (Figs.\ 
8 and 9). This implies that ${\rm (i)}$ the effective radial wavenumbers for these components are 
slightly different which can explain the deviation from the WKB result seen in Fig. 6, as
the velocity and magnetic field perturbations should have a common radial wavenumber in the local case
(although, formally the condition for the WKB approximation -- $r_o$ is larger than the radial lengthscale of 
the perturbation -- is satisfied) and ${\rm (ii)}$ the boundary conditions on the cylinders do play a role in 
determining the structure of the eigenfunctions and therefore can modify the growth rate of the instability. 
Indeed, it is seen in Figs. 8 and 9 that the insulating boundaries introduce more smaller scale features 
in the eigenfunctions, especially in the velocity ones, compared to those in the presence of the conducting boundaries, 
which apparently leads to the difference in the corresponding growth rates in Fig.\ 6. 

The situation is different in the case of large $Pm=100$ shown in Fig.\ 10 in that the axial 
lengthscale of the eigenfunctions is much smaller than the radial one, which is again 
comparable to the gap width, i.e., $k_{zm}\gg \lambda_r^{-1} \sim \delta^{-1}$ (for the ease of comparison, the same 
range in $z$ is used in all panels in Figs.\ 8-10). It is seen from this figure that unlike the 
case of small $Pm$, now the structure of the velocity and magnetic field eigenfunctions and hence 
their effective radial wavenumbers are very similar. Since this effective radial wavenumber turns 
out to be much smaller than $k_{zm}$, to a zero approximation the eigenfunctions can be assumed 
to be radially independent and vary only along $z$. In this case, however, calculating the growth 
rate, one arrives at the same WKB dispersion relation (Eq.\ 4) (when $k_r\gg k_z$, $\alpha\approx 
1$). This implies that actually the WKB regime is better fulfilled at large $Pm$ and those higher 
$k_z$, at which the instability reaches a maximum growth, and hence specific boundary conditions 
do not affect its growth and frequency. This can also explain the better agreement between 
the dispersion curves from the WKB and global analyses in Fig.\ 7 than that in Fig.\ 6. As a result, 
the structures of the eigenfunctions are also similar for both conducting and insulating boundaries. 
For this reason, we show only the case with the conducting boundaries in Fig.\ 10, leading to identical 
dispersion curves in Fig.\ 7. 

This comparative local and 1D global analysis of type 2 Super-HMRI, for the conducting and 
insulating boundary conditions, in the narrow gap case, allowed us, first of all, to clearly 
demonstrate the existence of this new double-diffusive instability also in the global setup of a differentially rotating 
dissipative flow with positive shear threaded by helical magnetic field. We have showed that the basic behavior 
of its growth rate and frequency on the flow parameters and axial wavenumber can still be understood and reproduced 
qualitatively within the local WKB approach. This indicates that type 2 Super-HMRI, like HMRI at 
negative shear, is a genuine instability intrinsic to the flow, tapping free energy of differential 
rotation (shear), and is ${\it not}$ induced/driven by the radial boundaries on the confining rotating 
cylinders, although it can still be modified by these boundaries. We have seen above that the 
quantitative differences between the results of the WKB and global analyses and some influence of 
specific radial boundary conditions on the growth rate and structure of the eigenfunctions are 
more noticeable at small magnetic Prandtl numbers, but diminish at large Prandtl numbers, 
because with increasing this number the eigenfunctions tend to vary much more along the 
axial $z$-direction than in the radial direction. Consequently, the WKB approximation holds much 
better in the latter case. 

\subsection{Wide-gap case}

Having explored type 2 Super-HMRI in the local and narrow-gap global cases, we 
now look for it in the case of a wide gap. As noted above, in this subsection, the primary goal of these calculations is 
to identify this new instability in a TC flow setup commonly employed in lab experiments, which 
in the present case has a radially increasing angular velocity (positive shear) profile. For 
this purpose, following \cite{Ruediger_etal2016,Ruediger_etal2018b,Ruediger_etal2018c}, where 
the related Super-AMRI is studied in a TC flow at positive shear, we adopt here one of the radial 
profiles for the angular velocity used in those papers, which is different from expression (6) at 
constant $Ro$. Specifically, the inner cylinder is assumed to be stationary, $\Omega_i=0$, while 
the outer cylinder rotates at $\Omega_o$, achieving a radially increasing angular velocity profile 
between the cylinders with the largest positive shear for a give rotation rate of the outer cylinder, 
\[
\Omega(r)=\frac{\Omega_o}{1-\hat{\eta}^2}\left(1-\frac{r_i^2}{r^2}\right),
\] 
where for the ratio of the inner to outer cylinders' radii we take the fixed value $\hat{\eta}=0.5$ relevant to the 
TC devices in the PROMISE and DRESDYN experiments \cite{Stefani_etal2006,Stefani_etal2019}. 
We take $\beta \sim 1$ in these calculations for the wide-gap, since it is rather costly to achieve 
large $\beta$ in experiments due to very high axial currents required. In this wide gap 
case, we impose conducting boundary conditions for the magnetic field on the cylinders. Note also 
that at $\beta=1$ the re-scaled parameters are equal to their actual counterparts: 
$k_z^{\ast}=k_z$, $Ha^{\ast}=Ha$, $Re^{\ast}=Re$, that is more convenient, allowing us to directly 
compare the results at wider gap with the local ones presented in Figs.\ 1 and 2.

\begin{figure}
\includegraphics[width=\columnwidth]{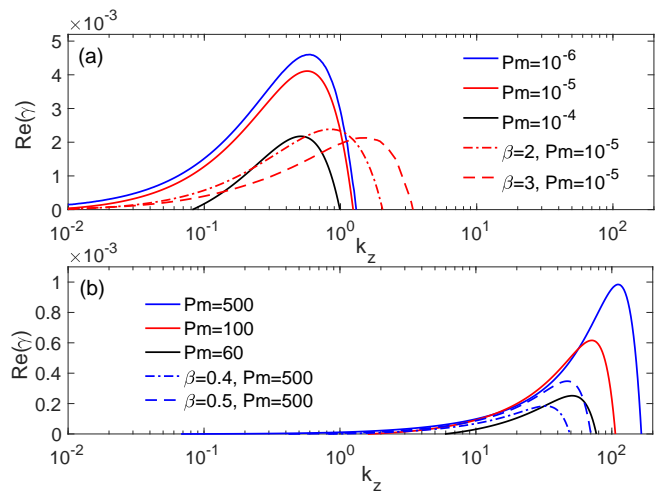}
\caption{Growth rate ${\rm Re}(\gamma)$ vs. $k_z$ for the TC flow with the wide gap, 
$\hat{\eta}=0.5$, and the inner cylinder at rest. Panel (a) focuses on $Pm\ll 1$, with solid lines 
having fixed $\beta=1$, $S=4$ and $Rm=20$ but different $Pm$, dot-dashed line $\beta=2$, 
$S=8$, $Rm=80$, $Pm=10^{-5}$, and dashed line $\beta=3$, $S=12$, $Rm=180$, 
$Pm=10^{-5}$. Panel (b) focuses on $Pm\gg 1$, with solid lines having fixed $\beta=1$, 
$Ha=1600$ and $Re=8\times10^4$ but different $Pm$, dot-dashed line $\beta=0.4$, $Ha=640$, 
$Re=1.28\times10^4$, $Pm=500$, and dashed line $\beta=0.5$, $Ha=800$, $Re=2\times10^4$, 
$Pm=500$.}
\end{figure}

\begin{figure}
\includegraphics[width=\columnwidth]{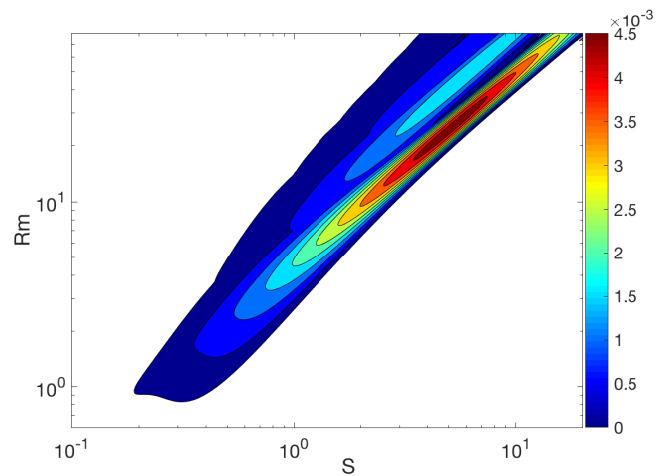}
\caption{Growth rate for the TC flow with the positive shear and wide gap, maximized over $k_z$ 
and represented in $(S,Rm)$-plane, at fixed $Pm=10^{-6}$ and $\beta=1$. As in the local 
equivalent (Fig.\ 1(c)), the unstable region is localized in $(S,Rm)$-plane, reaching a maximum at 
$S_m=5.2$ and $Rm_m=25$.}
\end{figure}

Figure 11 shows the growth rate ${\rm Re}(\gamma)$ as a function of the wavenumber $k_z$ for 
both small and large $Pm$, clearly demonstrating the presence of type 2 Super-HMRI also at a 
wide gap in the magnetic TC flow with radially increasing angular velocity. For $\beta=1$, the 
instability is concentrated mainly at small $k_z\lesssim 1$ for $Pm\ll 1$ and shifts to larger $k_z 
\gtrsim 10$ for $Pm\gg 1$ as it is in the local analysis (Figs.\ 1(a) and 2(a)). Note also that here the 
solutions explicitly depend on $\beta$, since the radial coordinate appears in the governing 
equations \cite{Hollerbach_Ruediger2005}, unlike in the local analysis where $\beta$ effectively 
scales out. However, as seen in Fig.\ 4, qualitatively the expected scalings with $\beta$ are still 
followed, with $Ha$ and $Re$ (or $S$ and $Rm$) as well as $k_{zm}$ increasing with $\beta$.

Note also in Fig.\ 11(a) how three solid curves, corresponding to small $Pm$, 
converge as $Pm$ decreases. This is consistent with the previous local analysis result that the relevant defining 
parameters in this case are Lundquist, $S$, and magnetic Reynolds, $Rm$, numbers, since they 
are kept fixed for these three curves. Figure 12 focuses further on this small magnetic Prandtl 
number case, as it is more relevant to TC experiments employing liquid metals (gallium, sodium) 
with $Pm\sim 10^{-6}-10^{-5}$ \cite{Stefani_etal2006,Stefani_etal2009, Seilmayer_etal2014}, and 
shows how the growth rate, maximized over $k_z$, varies with $S$ and $Rm$ at 
fixed $\beta=1$. Just as in Fig.\ 1(c) of the local analysis, the instability is again localized in 
$(S,Rm)$-plane, with the overall maximum growth rate, $\gamma_m=4.8\times 10^{-3}$, occurring 
at $S_m=5.2$ and $Rm_m=25$ ($Ha_m=5.2\times10^3$ and $Re_m=2.5\times 10^7$). The 
minimum, or critical values for the instability to first emerge are $S_c\approx 0.3$ and 
$Rm_c\approx 0.9$. These values are well within the capabilities of the new TC device at HZDR 
\cite{Stefani_etal2019}, offering a realistic prospect for experimental realization/detection of type 2 
Super-HMRI. However, since at $\beta \sim 1$ type 2 Super-HMRI occurs at smaller $k_z$ 
(Figs.\ 11(a)), as far as the experimental detection is concerned, optimization with respect 
to $\beta$ is required in order to ensure that the largest critical $k_{z, max}$ of the 
instability, corresponding to the smallest wavenumber $\lambda_{z,min}=2\pi/k_{z,max}$, still fits 
into the device, that is, $\lambda_{z,min}\leq L_z$, where $L_z$ is its vertical size. Such an optimization 
procedure and a more detailed analysis addressing the experimental manifestation of this 
instability will be presented elsewhere. 

Although the results at wider gap are qualitatively similar to local ones, there are noticeable quantitative 
differences in that the local analysis yields about an order of magnitude higher growth 
rate, but lower axial wavenumber and Hartmann number and about two orders of magnitude lower 
Reynolds number for the instability. This is mismatch is expected, 
since in this case of $\beta \sim 1$, as mentioned above, the WKB 
approximation is inapplicable. The radial and axial lengthscales of the eigenfunctions are 
comparable to the gap width as well as to the radial size of the flow system, $\lambda_z\sim \lambda_r\sim \delta = 
r_i =0.5r_o$) and therefore the equilibrium appreciably varies over the whole radial extent of the 
mode. By contrast, the local WKB analysis better applies in the narrow gap case, $\hat{\eta}\rightarrow1$, 
because the eigenfunctions vary mostly only over the gap width (Figs.\ 8-10), 
which is much less than $r_o$ and hence the radial variation of the equilibrium quantities are small across this 
distance. Another reason for the differences in the growth rates between the local and the global 
wide gap cases is that a finite distance between the cylinders also excludes (cuts off) very small 
radial wavenumbers ($\alpha\rightarrow1$), which correspond to larger growth rates at $\beta \sim 
1$ in the WKB analysis (see Fig.\ 1(b)).

The primary purpose of these 1D global calculations for the narrow and wide gap, supplementing 
the local analysis, has been to demonstrate the existence of this novel double-diffusive type 2 
Super-HMRI at positive shear in the global setup too. A more comprehensive global linear analysis 
of this instability, exploring the dependence of its growth rate on the flow parameters 
($\hat{\eta},\beta, Re, Ha,Pm$), will be presented elsewhere. We also plan to explore the effects 
of boundary conditions, conducting vs. insulating, in more detail at wide gap $\hat{\eta}=0.5$, 
since, as we have seen in the narrow gap case, they can lead to quite different onset criteria and 
growth rates of type 2 Super-HMRI, as it generally happens for MRI-type instabilities in TC flows 
(see e.g., Refs. \cite{Ruediger_etal2018a,Ruediger_etal2018c}). Such a study will be important 
for setting up a series of new tailored TC experiments with radially increasing angular velocity 
at the DRESDYN facility, aiming at detecting type 2 Super-HMRI.

\begin{figure}
\includegraphics[width=\columnwidth]{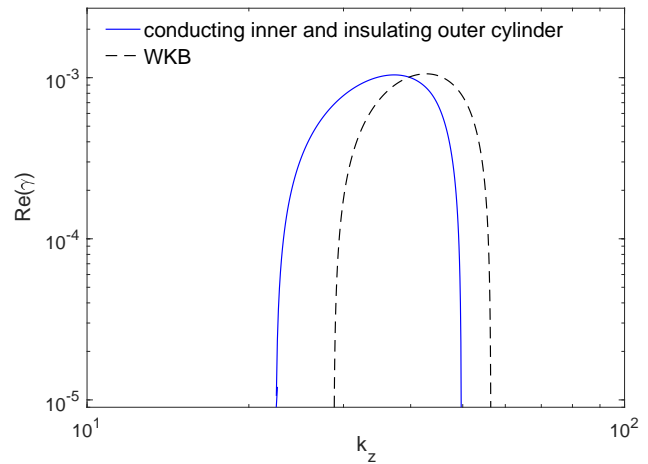}
\caption{Growth rate of type 2 Super-HMRI vs. $k_z$ for a power-law radial profile of the angular 
velocity with $Ro=0.7$ and a very narrow gap $\hat{\eta}=0.94$, which are close to those of the 
solar tachocline. The  boundary conditions are conducting at the inner and insulating at the outer 
walls, as often used in  solar tachocline studies. The other parameters are taken also of the same 
order as those for the tachocline (see text): $\beta=500$, $Pm=0.01$, $Ha=2500$ and 
$Re=1.5\times 10^6$. As in Fig.\ 6, black dashed line shows the WKB result for these 
parameters, but at much higher radial wavenumber $k_{r0}=\pi(1-\hat{\eta})^{-1}\approx 52.4$, 
corresponding to a narrower gap. }
\end{figure}

\section{Summary and discussion}

In this paper, we have uncovered and analyzed a novel type of double-diffusive axisymmetric 
HMRI, labelled type 2 Super-HMRI, which exists in rotational flows with radially increasing angular 
velocity, or positive shear of arbitrary steepness threaded by a helical magnetic field -- a 
configuration where magnetorotational instabilities were previously unknown. The only 
prerequisites are that $Pm \ne 1$ and the imposed magnetic field includes {\it both} axial and 
azimuthal components. First we identified this instability using the radially local WKB analysis and 
subsequently ascertained its existence via 1D global linear stability analysis by solving a 
boundary value problem with conducting or insulating boundary conditions on the rotating 
cylindrical walls containing the flow. In the global setup, we separately considered the cases of 
narrow gap and large azimuthal field, in order to compare with the results of the WKB analysis, and a wide gap 
and moderate azimuthal field, which is more relevant to experiments. The comparative analysis in 
the narrow gap case indicate that although the global and local WKB results are qualitatively 
similar, there are quantitative differences in the growth rates and associated axial wavenumbers 
mostly at small magnetic Prandtl numbers, since the radial and axial lengthscales of the modes are 
comparable to each other and to the gap width and hence boundary conditions appear to modify 
the growth rate. By contrast, in the case of large Prandtl numbers, axial lengthscale of the mode is much 
smaller than the radial one (which is again comparable to the gap width) and consequently the 
influence of the boundaries are not important and the agreement with the WKB analysis is much 
better. In any case, the radial extent of the mode is always comparable to the gap width between 
the cylinders, i.e., it is concentrated within the bulk of the flow and not near either of the 
boundaries, indicating that type 2 Super-HMRI is in fact intrinsic to the flow system and is not 
induced due to specific boundary conditions. In this way, the global stability analysis also confirms 
that this new instability is a real one and not just an artifact of the WKB approximation.

From an experimental perspective, in the wide gap case, we have also demonstrated the presence 
of type 2 Super-HMRI in a magnetized viscous and resistive TC flow with positive shear, helical 
magnetic field and such values of $\hat{\eta}, \beta, Ha, Re, Pm$, which can be well achieved 
in liquid metal lab. This promising result of the global stability analysis can, 
in turn, be a basis for future efforts aiming at the detection of this instability and thereby providing an 
experimental evidence for its existence. This will also allow us to make comparison with theoretical 
results. Building on the findings of this first study, future work will explore in greater detail 
the parameter space and the effect of boundary conditions on type 2 Super-HMRI, which are more 
relevant to those present in TC flow experiments. 

\subsection*{Applicability to the solar tachocline}

MRI has already been discussed in relation to the solar tachocline in several studies \cite{Ogilvie2007,Parfrey_Menou2007,Masada2011,Kagan_Wheeler2014,Gilman2018a}, showing that it 
can arise at middle and high latitudes, where the shear of the differential rotation is negative. The resulting 
small-scale MRI-turbulence is thought to prevent coherent magnetic dynamo action in these regions of the tachocline, 
thus explaining the rare occurrence or absence of the sunspots these latitudes. 

This new double-diffusive type 2 Super-HMRI, 
on the other hand, may arise and potentially have important implications for the dynamics and magnetic activity of the low latitude, near-equatorial region of the solar tachocline, since  necessary conditions for the development of this instability can be realized there. Indeed, in this part of the tachocline, we have: {\rm (i)} the positive radial shear of differential 
rotation with $Ro\sim 0.1-1$ \cite{Parfrey_Menou2007, Tobias_etal2007},  {\rm  (ii)} small 
magnetic Prandtl numbers $Pm\sim 0.01$ \cite{Arlt_etal2006, Brun_Zahn2006, Garaud2007},  
{\rm (iii)} very high Reynolds numbers (inverse of often used Ekman number), 
$Re\sim 10^{11}-10^{14}$ \cite{Brun_Zahn2006,Tobias_etal2007,Arlt2009}, which are, of course, too challenging to use in theoretical/numerical studies 
so lower values  $Re\gtrsim 10^6$ are usually adopted instead (e.g., \cite{Garaud2007, 
Garaud_Garaud2008} and {\rm (iv)} helical magnetic field. Regarding the magnetic 
field, MHD models of the solar tachocline (e.g., \cite{Ruediger_Kitchatinov1997, Ruediger_Kitchatinov2007,Garaud2007} and 
references therein) also indicate that poloidal magnetic field in the tachocline 
should be relatively large in order to  produce a thin tachocline layer, and therefore have large 
Hartmann numbers $Ha\sim 10^3$, while the toroidal magnetic field should be even stronger, 
several orders of magnitude higher than the poloidal one, resulting in very large $\beta$ values. In 
practice, theoretical/numerical studies often use $\beta\sim 10^2-10^3$ 
(e.g., \cite{Arlt_etal2007a, Arlt_etal2007b}), which in our case would ensure that the mode is 
confined within a fairly thin layer of the tachocline, which has a typical ratio of the inner to outer 
radii $\hat{\eta}\sim 0.9-0.97$ \cite{Antia_Basu2011}. 

To demonstrate that the type 2 Super-HMRI can indeed arise in the tachocline-like configuration 
described above and thus to get a flavor of its amplification efficiency and characteristic 
wavenumber in this case, in Fig.\ 13 we show the growth rate as a function of $k_z$ obtained from an analogous 
1D global stability analysis as in the narrow-gap case of Sec. IV, assuming a power-law radial 
profile for the angular velocity (Eq. 6) with smaller $Ro=0.7$, narrower gap $\hat{\eta}=0.94$ and 
the values of the remaining parameters also of the same order as those quoted above for the 
tachocline. As distinct from the above analysis, now we impose the boundary conditions often used 
in tachocline studies, namely conducting at the inner and insulating at the outer radius, mimicking, respectively,  
the conducting radiative interior and insulating convection zone in the Sun (e.g., 
\cite{Ruediger_Kitchatinov1997, Garaud_Garaud2008, Acevedo2013}. It seen from 
this figure that the largest growth rate is $\sim 10^{-3}$, or in dimensional variables $\sim 
10^{-3}\Omega_o$, and therefore the corresponding growth time is $\sim 10^3\Omega_o^{-1}$. 
Taking into account that the Sun's angular velocity of rotation in the near-equatorial region of the 
tachocline is about $\Omega_o\approx 2\pi\times445~ nHz=88.2~ yr^{-1}$ \cite{Howe_etal2000}, 
for the characteristic growth time of this instability we get $\sim 11~yr$, which is surprisingly close 
to the solar cycle period. This suggests that type 2 Super-HMRI can be quite relevant to the 
magnetic activity in the tachocline, and generally in the Sun. 

Although these calculations regarding the applications to the solar tachocline, are 
simplistic, still remaining within the cylindrical rotational flow, they are nevertheless encouraging 
and motivate us to investigate in more detail the possible role of type 2 Super-HMRI in the 
dynamics, transport processes and dynamo action in the tachocline using its more realistic model 
involving spherical (shell) geometry, radially and meridionaly varying angular velocity (those parts 
where shear is positive, $Ro>0$) and magnetic field  distribution and boundary conditions. Such a 
comprehensive study will allow us also to characterize in depth potential implications of type 2 
Super-HMRI in the dynamical processes in the tachocline. In particular, this instability could also 
resurrect the idea of a subcritical solar dynamo. Its axisymmetric ($m=0$) nature can help to 
overcome the difficulties that have been identified \cite{Zahn_etal2007} in getting the so-called 
Tayler-Spruit dynamo \cite{Spruit2002} to form a closed dynamo loop from the combination of the 
non-axisymmetric ($m=1$) Tayler instability and the $m=0$ $\Omega$-effect. Finding out whether this 
scenario is actually realized requires a further dedicated study using these more realistic tachocline 
models and is a subject of future work. In this connection, we would like to mention that an analogous mechanism for 
explaining solar magnetic cycles was put forward in \cite{Rogers2011}, where the combined effect of differential rotation and axisymmetric current-driven instability can cause equatorward propagating reversals of the large-scale azimuthal field in the tachocline and convection zone, as it is observed in the Sun, without relying on a classical dynamo ($\alpha$)-effect.

\begin{acknowledgments}
This project has received funding from the European Union's Horizon 2020 research and innovation 
programme under the Marie Sk{\l}odowska -- Curie Grant Agreement No. 795158 and the ERC 
Advanced Grant Agreement No. 787544 as well as from the Shota Rustaveli National Science 
Foundation of Georgia (SRNSFG, grant No. FR17-107). GM acknowledges support from the 
Alexander von Humboldt Foundation (Germany). We thank both anonymous referees for constructive criticism which has led to an extended version of this paper and improved the presentation of our results.
\end{acknowledgments}

\bibliography{biblio.bib}

\end{document}